\documentclass[12pt,a4paper]{article}

\usepackage{a4wide}
\usepackage{amsmath,amssymb,amsfonts,amsxtra,simpler-wick,mathrsfs} 
\usepackage{hyperref}
\usepackage{physics}
\usepackage{caption}
\usepackage{subcaption}
\usepackage[nosort,noadjust]{cite}
\usepackage[noabbrev]{cleveref}
\allowdisplaybreaks
\usepackage{braket}
\usepackage{placeins}
\usepackage{ifthen}
\usepackage{xifthen}
\usepackage[nosort,noadjust]{cite}
\newcommand*\diff{\mathop{}\!\mathrm{d}}
\newcommand*\Diff{\mathop{}\!\mathrm{D}}
\newcommand{\Ji}{J_{i_1\dots i_{q/2};j_1\dots j_{q/2}}}

\numberwithin{equation}{section}
\numberwithin{figure}{section}

\newcommand{\del}[0] {\partial}

\newcommand{\mycite}[3][]{[#3 \citenum{#2}\ifthenelse{\isempty{#1}}{}{, #1}]}

\newcommand{\RNum}[1]{\uppercase\expandafter{\romannumeral #1\relax}}

\begin{document}
\thispagestyle{empty}
\vspace*{-2.5cm}
\begin{center}
\hfill BONN--TH--2024--19
\end{center}
\vskip 0.6in
\begin{center}
{\bf \Large Measurement and Teleportation in the cSYK/JT Correspondence}
\end{center}  

\vspace*{1cm}

\centerline{Raphael Brinster$^a$, Stefan F{\"o}rste$^b$, Yannic Kruse$^b$, Saurabh Natu$^b$ }
\vskip 0.25in
\begin{center}{\it $^a$ Institut für Theoretische Physik III, Heinrich Heine University D{\"u}sseldorf, Universit{\"a}tsstraße 1, D-40225 D{\"u}sseldorf, Germany}
\vskip 0.1in
{\it $^b$ Bethe Center for Theoretical Physics, Physikalisches Institut der Universität Bonn,\\
Nussallee 12, D-53115 Bonn, Germany}
\vskip 0.15in
{\tt Raphael.Brinster@hhu.de}\\
{\tt forste@th.physik.uni-bonn.de}\\
{\tt ykruse@uni-bonn.de}\\
{\tt snatu@uni-bonn.de}
\end{center}
\vskip 0.15in
\begin{abstract}
 We consider two complex SYK models entangled in a thermofield double (TFD) state and investigate the effect of one-sided projective measurements. As measurement operator we choose single site charge operators. Performing a measurement results in a non-zero $U(1)$ charge. The entropy curve differs from the previously studied SYK model due to a thermodynamic phase transition that takes place after a certain charge is reached. We also match our results to a dual bulk description. Finally, a teleportation protocol is provided to support the notion of a traversable wormhole being formed.
\end{abstract}

\newpage

\tableofcontents
\newpage


\section{Introduction}
Recently, the effect of one-sided measurements on the SYK thermofield double (TFD) and its holographic dual was investigated in \cite{Antonini_2023} (see also \cite{Antonini_2022} for a more general setting). There, it was shown that after a critical number of fermions had been measured on one side, that side's information was teleported to the other side of the TFD. 
For the dual theory, a two sided black hole of JT gravity was considered. The two sides of the geometry correspond to two entangled copies of the SYK in a TFD state. It was argued that the extent of the entanglement wedge belonging to the right SYK copy depends on the number of measured sites on the left. For only few measured sites it is limited by the black hole horizon. Whereas, it extends all the way to the left side if the number of measured sites exceeds a critical value. Resulting curves for the R\'{e}nyi-2 entropy in an SYK computation could be matched to the von Neumann entropy of the bulk. By devising a teleportation protocol the authors of \cite{Antonini_2023} were able to support the picture that a traversable wormhole forms \cite{Maldacena_2013,
maldacena2018eternaltraversablewormhole}. 

In the present paper, we will study the effects of performing  multiple single site charge measurements on one side of the TFD state in the cSYK model.\footnote{Here, cSYK stands for the complex SYK model in which Majorana fermions have been replaced by Dirac fermions \cite{Gu_2020, Davison_2017}. The cSYK has a global $U(1)$ symmetry.}
We consider an example in which each measurement returns a positive charge eigenvalue. 
This produces an entropy curve that differs non-trivially from the real SYK result. We also provide a dual description matching the CFT computation. Finally, we devise a teleportation protocol suggesting the formation of a traversable wormhole in the gravity dual.

The text at hand will be structured as follows. In this section, we first want to reiterate the main results of \cite{Antonini_2023} and lay down the necessary groundwork for our own calculations. We will do so by reviewing relevant results of JT/SYK, but will swiftly refer the reader to the existing more in depth reviews. In \cref{section_syk_measurement}, we will investigate the effect of a one-sided measurement on the cSYK TFD, a pure state that is comprised of two entangled but non-interacting copies of the cSYK. In \cref{sec:bulk}, we consider the effects this measurement procedure has on the bulk geometry. We see that the bulk geometry is that of an eternal black hole which is cut off by an end of the world brane after measurement. In \cref{sec:teleportation}, we will conclude the preceding section by adopting a teleportation protocol. This will serve to prove that the measurement allows us to send information from one side of the TFD to the other. This in turn, corresponds to the wormhole in the bulk becoming traversable.

\subsection{Measurement in the real SYK TFD}
The SYK model describes an ensemble of $N$-Majorana fermions $\psi_i$ with random all-to-all coupling of $q$ particles. 
In the low energy and large $N$ limit, a subsector (described by the Schwarzian) is dual to two dimensional Jackiw-Teitelboim (JT) gravity \cite{Sachdev_1993, Kitaev_lec, Kitaev:2017awl, Maldacena_2016,Teitelboim:1983ux,Jackiw:1984je, Almheiri:2014cka,Jensen:2016pah,Engelsoy:2016xyb, Cvetic:2016eiv,Gross:2017hcz} (see also  \cite{Sarosi_2018, Rosenhaus:2018dtp, Mertens:2022irh,Turiaci:2024cad} for reviews).
The model is defined by the Hamiltonian 
\begin{equation}
    H = i^{q/2} \sum_{i_1 < \dots < i_N}^{N} J_{i_1, \dots, i_q} \psi_{i_1}\dots \psi_{i_q}.
\end{equation}
Here, the coupling constants $ J_{i_1, \dots, i_q}$ are drawn from a Gaussian distribution with zero mean. We typically average the partition function over the $ J_{i_1, \dots, i_q}$, which gives rise to an effective action. In the large $N$ limit, the system becomes nearly classical, so that it is well described by its saddle point approximation. We can thus capture the averaged system's dynamics simply by solving the equations of motion stemming from the effective action. We will demonstrate this later for the cSYK model.

In \cite{Antonini_2023}, the authors use the SYK model to construct a thermofield double (TFD), which they subsequently perform projective measurements on. A TFD state is comprised of two identical copies of the same system (usually referred to as ``left'' and ``right'' ) that are non-interacting and entangled. If a fermion on one side of the TFD is measured, the state is projected onto a less entangled state. Nevertheless, the two sides always stay entangled after measurement unless all fermions are measured, which the authors show by computing the mutual information as a function of the number of measured fermions $M$. The measurement operator they considered is the single fermion parity operator.

Upon measurement of a number of fermions on the left side, boundary conditions are imposed onto the system, which manifest themselves as an end of the world (ETW) brane in the bulk anchored on the left boundary \cite{cavalcanti2019studiesboundaryentropyadsbcft,kim2023endworldperspectivebcft,Suzuki_2022,Fujita_2011,Takayanagi_2011,Kourkoulou:2017zaj}. The ETW brane renders part of the bulk inaccessible to the bulk matter dual to the measured fermions. The bulk entropy is then calculated. The entanglement wedges of the two sides will initially stay more or less unchanged for a small number of fermions measured. Once a certain threshold is crossed, the quantum extremal surface associated with the right side will abruptly transition to extend almost to the ETW brane on the left side. Hence, as seen from the entanglement wedge transition the information about the bulk initially stored in the left side of the TFD is teleported to the right when the critical value is reached. A teleportation protocol was devised to corroborate that idea. 

In the work at hand, we will keep the general idea of one sided measurement of a TFD state but will replace the real SYK by its complex version and the Fermi parity by a $U(1)$-charge measurement. We will next introduce the complex SYK model. 

\subsection{The cSYK model} \label{sec:cSYK}
In this section, we give a brief review of the complex Sachdev-Ye-Kitaev model (cSYK). For a more comprehensive discussion see in particular \cite{Gu_2020,Davison_2017,Bhattacharya_2017}. 

We obtain the complex SYK model by replacing the Majorana fermions in the vanilla SYK model with Dirac fermions and imposing a global $U(1)$-symmetry of the action. The corresponding Hamiltonian is given by 
\begin{align}\label{cSYK_hamiltonian}
    H = \frac{1}{\left(2N^{\frac{q-1}{2}}\right)} \sum_{i_1,\dots,i_{q/2}, j_1, \dots, j_{q/2}}^N \Ji \, c_{i_1}^\dagger\dots c_{i_{q/2}}^\dagger c_{j_1}\dots c_{j_{q/2}} -\mu \sum_{i}^N c^\dagger_i c_i,
\end{align}
where the fermionic operators $c_i$, $c_i^\dagger$ obey the Clifford algebra 
\begin{align}
    \poissonbracket{c_i}{c_j^\dagger} = \delta_{ij}. \label{eq:clifford}
\end{align}
The random couplings $\Ji$ are drawn from a Gaussian distribution with
\begin{align}
		&\overline{\Ji} = 0,&
		\overline{\abs{\Ji}^2} &= J^2 \label{mean_variance_cSYK}
\end{align}
and satisfy the following relations,
\begin{align}
    J_{\dots ij \dots;\dots kl\dots} = - J_{\dots ji\dots;\dots kl\dots} = - J_{\dots ij\dots;\dots lk\dots } = J_{\dots kl\dots ;\dots ij\dots}^*.
\end{align}
The parameter $\mu$ is the chemical potential conjugate to the total charge.
The partition function $Z$ should be averaged over the couplings, which we shall indicate by an overline, as such 
\begin{align}
    \overline{Z} &= \overline{\iint \Diff c_i \Diff c_i^\dagger e^{-I}} \nonumber\\
    &= \iiint dJ \Diff c_i \Diff c_i^\dagger e^{-I-\frac{1}{2J^2}\sum^N_{i_1, \dots ,i_{N/2}, j_1, \dots , j_{N/2}}\abs{J_{i_1,\dots,i_{N/2}, j_1,\dots, j_{N/2}}}^2} \nonumber\\
    &= \iint \Diff c_i \Diff c_i^\dagger e^{-I_\text{eff.}}.
\end{align} By doing so and taking the large $N$-limit we obtain the effective action
\begin{align}\label{average_cSYK}
    I_\text{eff.} = \sum_i \int_0^\beta \dd \tau \;  c_i^\dagger \left( \del_\tau - \mu \right)c_i - \frac{J^2}{4N^3} \int_0^\beta \int_0^\beta \dd \tau_1 \dd \tau_2 \; \abs{\sum_i c_i^\dagger(\tau_1) c_i(\tau_2)}^4.
\end{align}
We can then go to a collective field version of this action, by re-expressing everything in terms of the propagator 
\begin{align}
G(\tau_1,\tau_2)\equiv \frac{1}{N} \sum_i^N c_i^\dagger\left(\tau_1\right) c_i \left(\tau_2\right) 
\end{align}
and introducing the self energy $\Sigma(\tau_1,\tau_2)$ as a Lagrange multiplier by adding to the action
\begin{align}
    -\frac{1}{2}\iint \diff \tau_1 \diff \tau_2  \Sigma(\tau_1,\tau_2) \left[ G(\tau_2,\tau_1) - \frac{1}{N} \sum_i^N c_i^\dagger\left(\tau_1\right) c_i \left(\tau_2\right) \right].
\end{align}
This yields (after integrating over the Dirac fermions) 
\begin{equation}
\begin{split}
    \frac{I_\text{eff.}}{N} = &- \log \det \left[ \del_\tau - \mu - \Sigma \right] 
    + \int_0^\beta \int_0^\beta \dd \tau_1 \dd \tau_2 \Sigma(\tau_1,\tau_2) G(\tau_2,\tau_1) 
    \\&
    - \frac{J^2}{4} \int_0^\beta \int_0^\beta \dd \tau_1 \dd \tau_2 G^{\frac{q}{2}}(\tau_1,\tau_2) G^{\frac{q}{2}}(\tau_2,\tau_1). \label{eq:eff_action}
\end{split}
\end{equation}
The Schwinger-Dyson equations for the collective fields $G(\tau_1,\tau_2)$ and $\Sigma(\tau_1,\tau_2)$ read 
\begin{align}
    &G = \left[ \del_\tau -\mu - \Sigma \right]^{-1}, \label{schwinger_dyson_G_cSYK} \\
    &\Sigma(\tau_1,\tau_2) = J^2 G^{\frac{q}{2}}(\tau_1,\tau_2) \left(-G(\tau_2,\tau_1)\right)^{\frac{q}{2}-1}. \label{schwinger_dyson_sigma_cSYK}
\end{align}
Finally, at low temperature it can be shown that \eqref{eq:eff_action} takes the form \cite{Davison_2017}
\begin{align}\label{schwarzian_cSYK}
\begin{split}
\frac{I_{\text{cSYK,eff}}}{N} =  &\int_0^\beta \dd \tau \; \frac{K}{2} \left( \del_\tau \varphi + i\left(\frac{2\pi \mathcal{E}}{\beta}\right) \del_\tau \epsilon \right)^2 \\
&- \frac{\gamma}{4\pi^2} \;\text{Sch}\left\{ \tan \left[ \frac{\pi}{\beta} (\tau + \epsilon(\tau)) \right],\tau \right\}.
\end{split}
\end{align}
We see that in contrast to the real SYK model, an additional term proportional to the charge compressibility $K$ appears. Here, diffeomorphisms are parametrised as $\tau + \epsilon(\tau)$ and $\varphi(\tau)$ is an additional phase field that results from $U(1)$ symmetry transformations of $G$ and $\Sigma$.
We can express the charge compressibility $K$, specific heat $\gamma$, chemical potential $\mu$ and the spectral asymmetry factor $\mathcal{E}$ (where $\mathcal{S}_0$ is the zero temperature entropy) \cite{Gu_2020,Davison_2017,Gaikwad_2020} as
\begin{align}
    &K^{-1} = \left(\frac{\partial^2 F}{\partial \mathcal{Q}^2}\right)_T\label{compressibility},\\
    &\gamma = -\left(\frac{\partial^2 F}{\partial T^2}\right)_{\mathcal{Q}},  \label{gamma}
    \\& 
    \mathcal{J}^2 = \frac{q^2 J^2}{ 2 \left(2 + 2\cosh{\mu\beta}\right)^{\frac{q}{2}-1}},
    \\& 
    \mu = \left(\frac{\partial F}{\partial \mathcal{Q}}\right)_T, 
    \\&
    \mathcal{E} = \frac{1}{2\pi} \left(\frac{\partial \mathcal{S}_0}{\partial\mathcal{Q}}\right)\label{curlyE}.
\end{align}
Here, $\mathcal{Q}$ is defined as the charge density associated with the global $U(1)$-symmetry $c_k \mapsto e^{i\varphi} c_k$
\begin{align}
    \mathcal{Q} = \frac{1}{N} \sum_k \expval{Q_k}, & -\frac{1}{2} \leq \mathcal{Q} \leq \frac{1}{2},
\end{align}
where $Q_k$ is the charge operator at site $k$, i.e.\
\begin{equation}\label{cSYK_charge_operator}
    Q_k =  c_k^\dagger c_k - \frac{1}{2}.
\end{equation}

\subsection{Holographic dual of the cSYK model}\label{theory_dual_cSYK}
The action that has been proposed in \cite{Gaikwad_2020} as the holographic dual of the complex SYK model consists of JT gravity plus a Kaluza-Klein reduced $U(1)$ Chern-Simons field with a coupling term (the topological Einstein-Hilbert term is neglected here, it is just a constant that plays the role of the ground state entropy, same as in pure JT-gravity)
\begin{align}\label{action_JT+CS}
    I_{\text{JT+gauge}} =&- \frac{1}{16\pi G_2} \int_{\mathcal{M}} \sqrt{h} \phi \left(R + \frac{2}{l^2} - \frac{l^2}{4} \phi^2 \tilde{F}^2  \right) - \frac{ikl}{2} \int_{\mathcal{M}} \sqrt{h} \, \chi \left( J_0 \phi - F \right) \nonumber \\
    &+\frac{ikl}{4} \int_{\del \mathcal{M}}  \chi_b A_b - \frac{kl}{8} \int_{\del \mathcal{M}} \sqrt{h} \, \phi_b \, h_b \left( A_b^2 + h_b \left( \frac{\chi_b}{\phi_b} \right)^2 + \left( l \, \chi_b B_b \right)^2 - 2l \, \chi_b B_b A_b \right) \nonumber\\
    &- \frac{1}{8 \pi G_2}  \int_{\del \mathcal{M}} \sqrt{h} \, \phi_b \left(K - \frac{1}{l}\right).
\end{align}
This action can be obtained via Kaluza-Klein reduction from a three dimensional gravity theory coupled to a $U(1)$ Chern-Simon field. Similar to the dual of the real SYK model, JT gravity appears (for a more detailed overview see \cite{Almheiri:2014cka,mertens2023solvable,maldacena2016conformalsymmetrybreakingdimensional,saad2019jt,stanford2019jt}), as well as additional terms. $B_b$ corresponds to a gauge vector field (with field strength $\Tilde{F}$) and $A_b$ to the KK-reduced $U(1)$ Chern-Simons field (with field strength $F$). $\phi_b$ and $\chi_b$ are scalar fields arising from the dimensional reduction. For a derivation of \eqref{action_JT+CS} see \cite{Gaikwad_2020, Sachdev_2019,Moitra:2018jqs}. In the low temperature limit, the action above reduces to
\begin{equation}\label{eq: cSYK dual effective action}
    S_{\text{eff}} = \int_0^{\beta}\dd u \; \phi_r \; \left(\frac{k}{16}(\varphi'(u)+B\epsilon'(u))^2 - \frac{1}{8\pi G_2}\;\text{Sch}\left\{\tan\left(\frac{\pi}{\beta}\epsilon(u)\right),u\right\}\right).  
\end{equation}
Here, the first term corresponds to the contribution from the gauge field while the second term is the Schwarzian contribution from pure JT theory. The constant\footnote{$B$ is the coordinate independent part of the gauge field $A_b$.} $B$ is dual to the chemical potential and is related to $\phi_r$ via
\begin{equation}
    B \propto -\frac{\phi_r l_{\text{AdS}}^2}{\beta^2}.
\end{equation}
By comparing the holographic bulk dual, \eqref{eq: cSYK dual effective action}, to the effective action of the cSYK, \eqref{schwarzian_cSYK}, the relation between all the coefficients can be established,
\begin{align}
    &\phi_r = \frac{1}{\mathcal{J}}, \label{id_phi_r}\\
    &k =8\mathcal{J}K, \label{id_K}\\
    & B = i \frac{2\pi \mathcal{E}}{\beta},\label{iden_B}\\ 
    & \frac{1}{G_2} = \frac{2\mathcal{J}}{\pi}\gamma.\label{ide_G_2}
\end{align}
We can interpret the spectral asymmetry parameter $\mathcal{E}$ as being the electric field on the black hole horizon in the bulk \cite{Davison_2017,Sachdev_2015}.

\section{Charge Measurement of the cSYK Thermofield Double}\label{section_syk_measurement}
In this section, we will introduce measurement on one side of the TFD state (which we will subsequently refer to as the left side) of the complex SYK model. More specifically, a $U(1)$-charge measurement on $M$ of the $N$ fermions will be performed. We calculate the entanglement entropy $S_L(m)$, with $m \equiv M/N$, which will be approximated by the R\'{e}nyi-2 entropy (we show that this is a valid approximation in \cref{sec:neumannvrenyi}). 
Specifically, we focus on a situation in which the outcome of each measurement is positive.
We observe, that the entropy decreases as more fermions are measured and diminishes completely, after a critical fraction of fermions are measured.

\subsection{Boundary conditions from measurement and Euclidean path integral}
We define the measurement operator (notice the different normalisation in contrast to (\ref{cSYK_charge_operator}))
\begin{equation}
    Q_k = 2 c_k^\dagger c_k - 1 \label{eq:Charge_at_site_k}.
\end{equation}
$Q_k$ measures the charge of the $k$'th fermion and has eigenvalues $\pm 1$.
\noindent
Upon charge measurement of a subset consisting of the first $M$ fermions, the corresponding post-measurement state $\ket{L_l(m)}$ will be an eigenstate of all the $M$ measured charges
\begin{equation}\label{Q_k_eigenstate}
    Q_k \ket{L_l(m)} = l_k \ket{L_l(m)},
\end{equation}
with $k = 1,..., M$ and $l_k = \pm 1$.
Notice that (\ref{Q_k_eigenstate}) implies
\begin{align}\label{state_boundary_conditions}
    \left(c_k^\dagger - c_k \right)|L_l(M)\rangle = l_k \left(c_k +c_k^\dagger \right) |L_l(M)\rangle. 
\end{align}
Equation (\ref{state_boundary_conditions}) will later be used to calculate the new boundary conditions for the measured fermions.

The thermofield double state in the cSYK model is defined as the pure state \cite{Sahoo_2020}
\begin{equation}\label{TFD_cSYK}
		|\text{TFD}\rangle = \frac{1}{\sqrt{Z_\beta}} \sum_{Q=-N}^{N} \sum_{n_Q} e^{-\frac{\beta}{2}( E_n -\mu Q)} |n_Q\rangle_L \otimes | \Theta n_Q\rangle_R \:= e^{-\frac{\beta}{4}(H_\text{L}+H_\text{R})}\ket{\infty},
\end{equation}
where $\ket{n_Q}$ are energy eigenstates with charge $Q$ and in the second line we defined the TFD state at infinite temperature $\ket{\infty}\equiv \ket{\text{TFD}}_{\beta = 0}$. $\Theta$ is an anti-unitary operator that leaves the Hamiltonian invariant (e.g.\ the CPT operator 
\cite{Sahoo_2020,hartman2015lectures}). 

Since the measurement projects the TFD-state onto eigenstates of the $Q_k$, the unnormalised post-measurement state has the following form
\begin{equation}
    \ket{\psi_l(M)} = \left[ \left( \ketbra{L_l(M)}{L_l(M)} \otimes \mathbb{I}_L^{N-M}\right) \otimes \mathbb{I}_R \right] \ket{\text{TFD}}.
\end{equation}
Here, $\mathbb{I}_R$ reflects the completely untouched right side of the TFD and $\mathbb{I}_L^{N-M}$ the $(N-M)$ unmeasured fermions on the left side. We denote the unnormalised density matrix of the complete post-measurement state of the first $M$ fermions by $\psi_l(M) = \ketbra{\psi_l(M)}{\psi_l(M)}$ and the mutual information of $\psi_l(M)$ is defined as
\begin{equation}
    I_{LR}\left[\psi_l(M)\right] = S_L\left[\psi_l(M)\right] + S_R\left[\psi_l(M)\right] - S_{LR}\left[\psi_l(M)\right].
\end{equation}
Here $S$ denotes the entanglement entropy, where $S_{L/R}$ stands for the entanglement entropy of the reduced systems, where either the right or left system is being traced out, e.g.
\begin{equation}
    S_R\left[\psi_l(M)\right] \equiv S\left[ \Tr_L{\psi_l(M)} \right].
\end{equation}
The entropy of the full system, $S_{LR}(\psi_l(M))$, vanishes ($\psi_l(M)$ is a pure state). For the same reason, $S_L\left[\psi_l(M)\right] = S_R\left[\psi_l(M)\right]$ and therefore we have
\begin{equation}
    I_{LR}\left[\psi_l(M)\right] = 2 S_R\left[\psi_l(M)\right].
\end{equation}
Thus, the mutual information just corresponds to twice the entanglement entropy of either the left or the right side of the measured TFD.

The R\'{e}nyi-$n$ entropy is defined as 
\begin{equation}\label{theory_renyi}
    S^{(n)}(\rho) =  \frac{1}{1-n} \log \Tr \rho^n .
\end{equation}
It can be computed by using the replica trick (see e.g.\ \cite{Callebaut_2023}).
Here, we will approximate the entanglement entropy by the R\'{e}nyi-2 entropy, i.e.\
\begin{equation}\label{renyi-2}
    e^{-S_R} = \frac{\Tr_R\left[ \left( \Tr_L \psi_l(M) \right)^2 \right]}{ \left(\Tr \psi_l(M) \right)^2}.
\end{equation}
The numerator comes directly from the definition of the R\'{e}nyi-2 entropy (\ref{theory_renyi}) and the denominator is the normalisation. Numerator and denominator of equation (\ref{renyi-2}) can be translated to Euclidean path integrals with corresponding boundary conditions. The projection operators set the boundary conditions for the path integral (see below).

Both path integrals (one for the numerator, one for the denominator) will be calculated in saddle point approximation.
The entanglement entropy can therefore be approximated as
\begin{align}\label{entropy_saddle_point}
	S_R \approx I^\ast_{\mathit{num}} - I^\ast_{\mathit{den}}, 
\end{align}
with the on shell action $I^\ast$.
Numerical calculations for this will be conducted in the next subsection.

\begin{figure}[h!]
    \includegraphics[width=15cm]{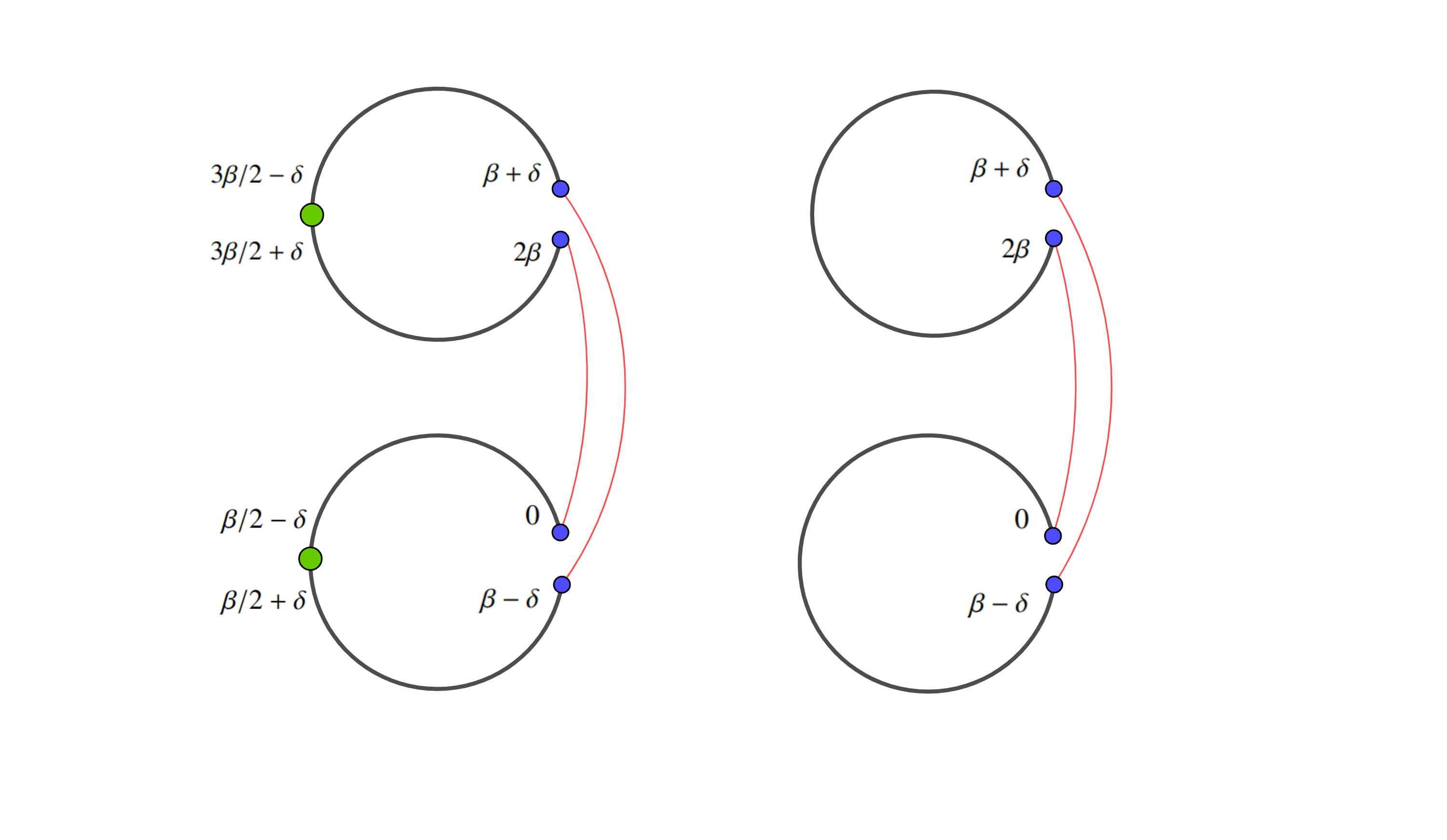}
    \centering
    \caption{Imaginary time contour of numerator path integral, left: measured fermions with measurement denoted by green dots and twist operator (connecting both replicas) by red line, setting anti-periodic b.c. with period $2\beta$; right: unmeasured fermions, no measurement inserted.}
    \label{fig:path_integral_numerator}
\end{figure}

To account for the replica arising in calculating the R\'{e}nyi-2 entropy, the imaginary time is extended to range from $0$ to $2\beta$, where $\tau \in (\beta,2\beta)$ belongs to the replica. 
We measure $Q_k$ at $\tau=\beta/2$ and $\tau = 3\beta/2$ in the replica. We select the cases in which $Q_k=1$ and therefore our measurement corresponds to
inserting the projection operators $\ketbra{L_1(m)}$. Additionally, for the numerator one has to insert a twist operator setting anti-periodic boundary conditions at $\tau = 0, 2\beta$ see figure \ref{fig:path_integral_numerator}.
\\
\\
First, we consider the numerator of \eqref{renyi-2}. In total, inserting the projection operators $\ketbra{L_1(M)}{L_1(M)}$ in the trace yields the following boundary conditions on the measured fields for the numerator ($k \in (1,..., M)$)
\begin{align}\label{bc_meas_num}
    -&\left( c_k^\dagger - c_k \right)\left(\frac{\beta_-}{2}\right) = \left( c_k^\dagger + c_k \right)\left(\frac{\beta_-}{2}\right), & & \left( c_k^\dagger - c_k \right)\left(\frac{\beta_+}{2}\right) =  \left( c_k^\dagger + c_k \right)\left(\frac{\beta_+}{2}\right), \nonumber \\
    -& \left( c_k^\dagger - c_k \right)\left(\frac{3\beta_-}{2}\right) =  \left( c_k^\dagger + c_k \right)\left(\frac{3\beta_-}{2}\right), & & \left( c_k^\dagger - c_k \right)\left(\frac{3\beta_+}{2}\right) = \left( c_k^\dagger + c_k \right)\left(\frac{3\beta_+}{2}\right), \nonumber \\
    &c_k(0) = - c_k(2\beta), & &  c_k(\beta_-) = c_k(\beta_+), \nonumber \\
    &c_k^\dagger(0) = - c_k^\dagger(2\beta), & &  c_k^\dagger(\beta_-) = c_k^\dagger(\beta_+).
\end{align}
Here, $\beta_{\pm} = \beta \pm \delta$, where $\delta$ is some small positive real number. The first two equations in (\ref{bc_meas_num}) arise due to the measurement. The easiest way to see this is from equation \eqref{state_boundary_conditions} and the path integral (\cref{fig:path_integral_numerator}). Noting, that the Euclidean path integral prepares the ket at $\tau = \frac{\beta_+}{2}$ and the bra at $\tau = \frac{\beta_-}{2}$ (or $\tau = \frac{3\beta_{\pm}}{2}$ for the second replica). The relative minus signs when comparing the equations for $\tau = \frac{\beta_+}{2}$ and $\tau = \frac{\beta_-}{2}$ come from taking the hermitian conjugate of equation (\ref{state_boundary_conditions}) when evaluating the equation for the bra. The last two equations in (\ref{bc_meas_num}) are due to the twist operator applied on the right side.

The unmeasured fermions only fulfil the usual anti-periodic boundary conditions for the replicated geometry coming from the twist operator
\begin{align}\label{bc_unmeas_num}
    &c_i(0) = - c_i(2\beta), & &  c_i(\beta_-) = c_i(\beta_+), \nonumber \\
    &c_i^\dagger(0) = - c_i^\dagger(2\beta), & &  c_i^\dagger(\beta_-) = c_i^\dagger(\beta_+).
\end{align}

\begin{figure}[h]
    \includegraphics[width=15cm]{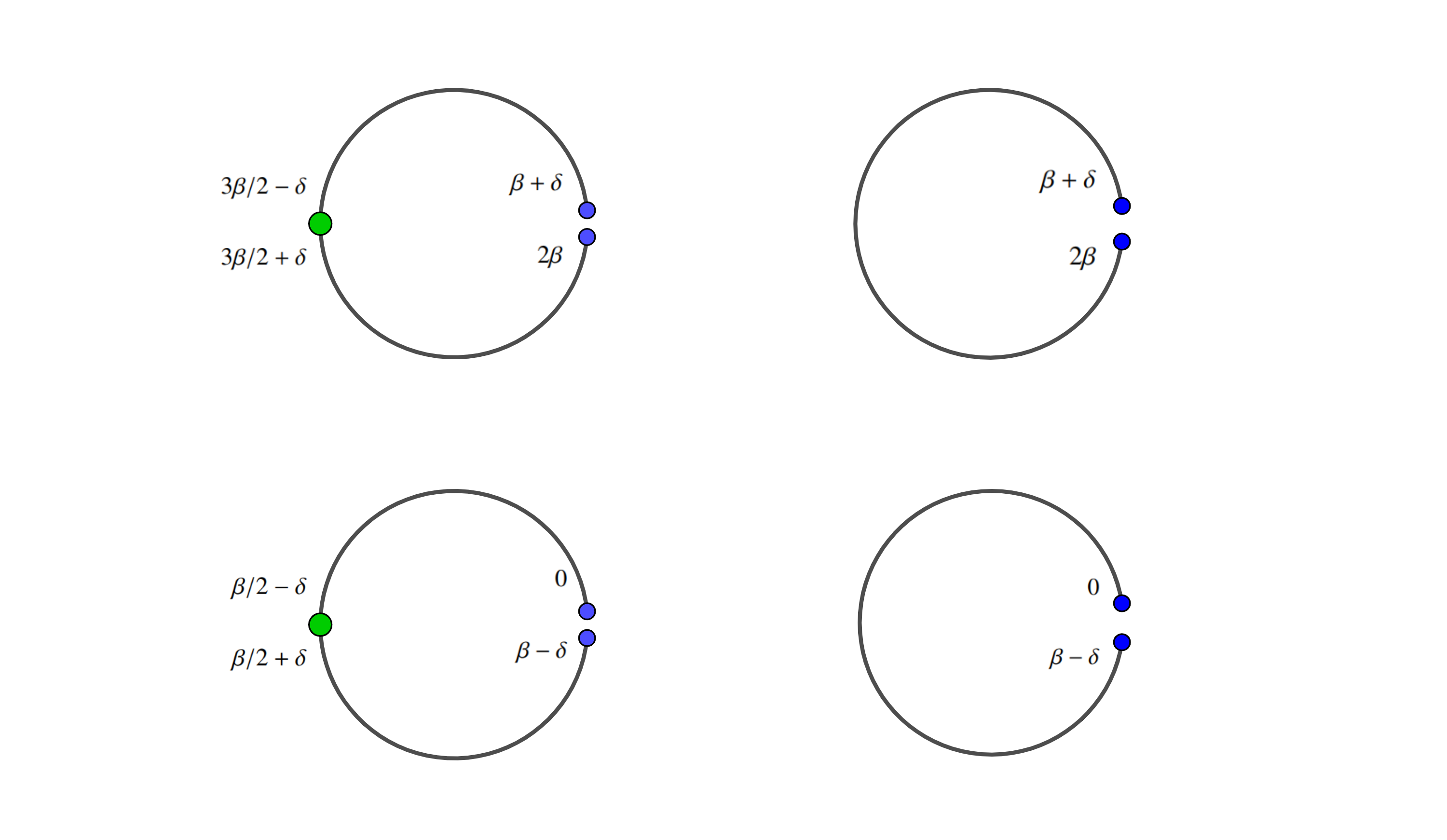}
    \centering
    \caption{Imaginary time contour of denominator path integral, left: measured fermions with measurement denoted by green dots and no twist operator; right: unmeasured fermions, no measurement inserted and no twist operator.}
    \label{fig:path_integral_denominator}
\end{figure}

Similar arguments can be made for the path integral in the denominator of equation (\ref{renyi-2}). The projection operators $\ketbra{L_1(M)}{L_1(M)}$ in the trace lead to the same boundary conditions at the measurement points $\tau = \frac{\beta_{\pm}}{2},\frac{3\beta_{\pm}}{2}$. But since we are not calculating $\Tr \rho_R^2$ as for the numerator but rather $(\Tr \rho)^2$, no twist operator is inserted. The fermions show anti-periodicity under shifts by 
$\beta$ on the two geometries separately (see figure \ref{fig:path_integral_denominator})
\begin{align}\label{bc_meas_den}
	-&\left( c_k^\dagger - c_k \right)\left(\frac{\beta_-}{2}\right) =  \left( c_k^\dagger + c_k \right)\left(\frac{\beta_-}{2}\right), & & \left( c_k^\dagger - c_k \right)\left(\frac{\beta_+}{2}\right) =  \left( c_k^\dagger + c_k \right)\left(\frac{\beta_+}{2}\right), \nonumber \\
	-& \left( c_k^\dagger - c_k \right)\left(\frac{3\beta_-}{2}\right) =  \left( c_k^\dagger + c_k \right)\left(\frac{3\beta_-}{2}\right), & & \left( c_k^\dagger - c_k \right)\left(\frac{3\beta_+}{2}\right) =  \left( c_k^\dagger + c_k \right)\left(\frac{3\beta_+}{2}\right), \nonumber \\
	&c_k(0) = - c_k(\beta_-), & &  c_k(\beta_+) = -c_k(2\beta), \nonumber \\
	&c_k^\dagger(0) = - c_k^\dagger(\beta_-), & &  c_k^\dagger(\beta_+) = -c_k^\dagger(2\beta).
\end{align}
Similarly for the unmeasured fermions
\begin{align}\label{bc_unmeas_den}
&c_i(0) = - c_i(\beta_-), & &  c_i(\beta_+) = - c_i(2\beta), \nonumber \\
&c_i^\dagger(0) = - c_i^\dagger(\beta_-), & &  c_i^\dagger(\beta_+) = - c_i^\dagger(2\beta).
\end{align}

\subsection{Large-{\textit{N}} action and Schwinger-Dyson equations}
In the following, we will derive the large-$N$ action of the cSYK model in terms of the bilocal fields $G(\tau_1,\tau_2)$ and $\Sigma\left(\tau_1,\tau_2 \right)$. However, before integrating out the fermions the boundary conditions derived in the last subsection need to be implemented. After that, the Schwinger-Dyson equations can be derived and then solved iteratively to evaluate the on-shell action needed for the R\'{e}nyi-2 entropy.

We start off with the disorder-averaged large-$N$ effective action (here, the Euclidean time ranges from $0$ to $2\beta$, since we are considering the replicated geometry),
\begin{align}\label{large_N_final_q}
    -I = -\int_{0}^{2\beta} \dd{\tau} c_i^\dagger \partial_\tau c_i
	    + \frac{J^2}{q N^{q-1}} \int_{0}^{2\beta} \int_{0}^{2\beta} \dd{\tau_1} \dd{\tau_2} \abs{ \sum_{i} c_i^\dagger(\tau_1) c_i(\tau_2) }^q
\end{align}
Due to the different boundary conditions for the measured and unmeasured fermions we should distinguish between them in the action. By re-expressing the sums in \eqref{large_N_final_q} in terms of $c_k^{\dagger} + c_k$ and $c_k^{\dagger}-c_k$ for the measured fields we get
\begin{samepage}
\begin{align}\label{large_N_sum_diff}
    -I = \iint \dd{\tau_1} \dd{\tau_2}& \left( \frac{1}{4} \left(c_k - c_k^\dagger \right)(\tau_1) \partial \left(c_k - c_k^\dagger \right)(\tau_2) - \frac{1}{4} \left(c_k + c_k^\dagger \right)(\tau_1) \partial \left(c_k + c_k^\dagger \right)(\tau_2) \right. \nonumber \\
    &-c_i^\dagger(\tau_1) \partial  c_i(\tau_2) \nonumber \\ 
    + \frac{J^2}{q N^{q-1}} &  \left| -\frac{1}{4} \left(c_k-c_k^\dagger \right)(\tau_1) \left(c_k-c_k^\dagger \right)(\tau_2) +\frac{1}{4} \left(c_k+c_k^\dagger \right)(\tau_1) \left(c_k+c_k^\dagger \right)(\tau_2) \right. \nonumber\\
    &  -\frac{1}{4} \left(c_k-c_k^\dagger \right)(\tau_1)\left(c_k+c_k^\dagger \right)(\tau_2) +\frac{1}{4} \left(c_k+c_k^\dagger \right)(\tau_1) \left(c_k-c_k^\dagger \right)(\tau_2) \nonumber \\
    & + \left. \left. c_i^\dagger(\tau_1) c_i(\tau_2) \right|^q \right),
\end{align}
\end{samepage}
where we use the sum convention and denote the measured fermions with subscript $k\in (1,\cdots, M)$ and the unmeasured ones with subscript $i\in (M+1,\cdots, N)$. Additionally, the abbreviation $\del \equiv \delta(\tau_1-\tau_2)\del_{\tau_2}$ is introduced.

Now, it is possible to substitute in the two-point-function and introduce the self-energies $\Sigma$ as Lagrange multipliers in the large-$N$ action. Same as in \cite{Antonini_2023} we get ``diagonal'' as well as ``off-diagonal'' propagators for the unmeasured fermions. Furthermore, off-diagonal contributions to the self-energy appear, which are absent in the real SYK model. The action takes the following form
\begin{align}\label{eq: Collective field rep}
    -I =  \iint \dd{\tau_1} \dd{\tau_2} &  + \frac{1}{4} \left(c_k - c_k^\dagger \right)(\tau_1)\left[ \partial + \Sigma_{11}(\tau_2,\tau_1) \right] \left(c_k - c_k^\dagger \right)(\tau_2) \nonumber \\
    &+ \frac{1}{4} \left(c_k - c_k^\dagger \right)(\tau_1)[ \Sigma_{12}(\tau_2,\tau_1)] \left(c_k + c_k^\dagger \right)(\tau_2) \nonumber \\ 
    &+ \frac{1}{4} \left(c_k + c_k^\dagger \right)(\tau_1)[\Sigma_{21}(\tau_2,\tau_1)] \left(c_k - c_k^\dagger \right)(\tau_2) \nonumber \\
    &- \frac{1}{4} \left(c_k + c_k^\dagger \right)(\tau_1)[\partial - \Sigma_{22}(\tau_2,\tau_1)] \left(c_k + c_k^\dagger \right)(\tau_2) \nonumber \\ 
    &- c_i^\dagger(\tau_1)[\partial - \Sigma_{33}(\tau_2,\tau_1)]c_i(\tau_2) \\
    & -\frac{M}{4} \Sigma_{11}(\tau_1,\tau_2) G_{11}(\tau_2,\tau_1) -\frac{M}{4} \Sigma_{12}(\tau_1,\tau_2) G_{12}(\tau_2,\tau_1) \nonumber \\
    &  - \frac{M}{4} \Sigma_{21}(\tau_1,\tau_2) G_{21}(\tau_2,\tau_1) - \frac{M}{4} \Sigma_{22}(\tau_1,\tau_2) G_{22}(\tau_2,\tau_1) \nonumber \\
    & - (N-M) \Sigma_{33}(\tau_1,\tau_2) G_{33}(\tau_2,\tau_1) \nonumber \\
    & \left. + \frac{J^2}{q N^{q-1}} \left(-\frac{M}{4} G_{11}(\tau_1,\tau_2) - \frac{M}{4} G_{12} + \frac{M}{4} G_{21} + \frac{M}{4} G_{22}  + (N-M)G_{33} \right)^{\frac{q}{2}} \right. \nonumber \\
    &\hphantom{+\frac{J^2}{q N^{q-1}}} \!\!\!\!\!\times  \left(-\frac{M}{4} G_{11}(\tau_2,\tau_1) - \frac{M}{4} G_{12} + \frac{M}{4} G_{21} + \frac{M}{4} G_{22}  + (N-M)G_{33} \right)^{\frac{q}{2}}. \nonumber 
\end{align}
Plugging in the definitions for the measured fields $G_{ab}(\tau_1,\tau_2)$ and $\Sigma_{ab}(\tau_1,\tau_2)$ with $a,b \in (1,2)$ and the unmeasured fields $G_{33}(\tau_1,\tau_2)$ and $\Sigma_{33}(\tau_1,\tau_2)$ set by the equations of motion
\begin{align}
    G_{11}(\tau_1,\tau_2) =& \frac{1}{M} \sum_{i=1}^{M} (c_k - c_k^\dagger)(\tau_1)(c_k - c_k^\dagger)(\tau_2) \label{eq:G11}\\
    G_{22}(\tau_1,\tau_2) =& \frac{1}{M} \sum_{i=1}^{M} (c_k + c_k^\dagger)(\tau_1)(c_k + c_k^\dagger)(\tau_2) \label{eq:G22}\\
    G_{12}(\tau_1,\tau_2) =& \frac{1}{M} \sum_{i=1}^{M} (c_k - c_k^\dagger)(\tau_1)(c_k + c_k^\dagger)(\tau_2)\label{eq:G12}\\
    G_{21}(\tau_1,\tau_2) =& \frac{1}{M} \sum_{i=1}^{M} (c_k + c_k^\dagger)(\tau_1)(c_k - c_k^\dagger)(\tau_2)\label{eq:G21}\\
    G_{33}(\tau_1,\tau_2) =& \frac{1}{N-M} \sum_{i=M+1}^{N} c_i^\dagger(\tau_1) c_i(\tau_2) \label{eq:G33}
\end{align}
we get back the action (\ref{large_N_sum_diff}) only in terms of the fields $c$ and $c^\dagger$.
The next step is to integrate out the fermionic fields $c$ and $c^\dagger$. To do so we first define a new fermionic field $\chi_k(s)$, which consists of the measured fields $c_k$ and $c_k^\dagger$
\begin{equation}\label{chi}
    \chi_k(s) = \frac{i}{\sqrt{2}}
    \begin{cases} 
    (c_k - c_k^\dagger) (s), & 0 < s < \frac{\beta}{2}, \\
    (c_k + c_k^\dagger) (\beta -s), & \frac{\beta}{2} < s < \beta, \\
    -(c_k + c_k^\dagger) (2\beta - s), & \beta < s < \frac{3\beta}{2}, \\
    (c_k - c_k^\dagger) (s - \beta), & \frac{3\beta}{2} < s < 2\beta, \\
    (c_k - c_k^\dagger) (s - \beta), & 2\beta < s < \frac{5\beta}{2}, \\
    (c_k + c_k^\dagger) (4\beta -s), & \frac{5\beta}{2} < s < 3\beta, \\
    -(c_k + c_k^\dagger) (5\beta - s), & 3\beta < s < \frac{7\beta}{2}, \\
    (c_k - c_k^\dagger) (s - 2\beta), & \frac{7\beta}{2} < s < 4\beta. 
    \end{cases}
\end{equation}
The field $\chi_k(s)$ is piecewise defined over the doubled range $s \in (0,4\beta)$. The boundary conditions of the measured fermions in (\ref{bc_meas_num}) or (\ref{bc_meas_den}) ensure continuity and anti-periodicity of the piecewise defined field. As mentioned before, we set all $l_k = 1$. This amounts to post-selecting the measurement outcome.
The different boundary conditions (numerator or denominator) will lead to different free propagators as initial input when iteratively solving the Schwinger-Dyson equations, see below.

In terms of the $\chi_k(s)$ field, the kinetic term can be re-expressed as follows
\begin{align}\label{kin_chi}
    \int_0^{2\beta} \int_0^{2\beta} \dd{\tau_1} \dd{\tau_2} & \left[ + \frac{1}{4} \left(c_k - c_k^\dagger \right)(\tau_1)\left[ \partial + \Sigma_{11}(\tau_2,\tau_1) \right] \left(c_k - c_k^\dagger \right)(\tau_2) \right. \nonumber \\
    &+ \frac{1}{4} \left(c_k - c_k^\dagger \right)(\tau_1)[ \Sigma_{12}(\tau_2,\tau_1)] \left(c_k + c_k^\dagger \right)(\tau_2) \nonumber \\ 
    &+ \frac{1}{4} \left(c_k + c_k^\dagger \right)(\tau_1)[\Sigma_{21}(\tau_2,\tau_1)] \left(c_k - c_k^\dagger \right)(\tau_2) \nonumber \\
    & \left. - \frac{1}{4} \left(c_k + c_k^\dagger \right)(\tau_1)[\partial - \Sigma_{22}(\tau_2,\tau_1)] \left(c_k + c_k^\dagger \right)(\tau_2) \right] \nonumber \\
    =&-\frac{1}{2} \int_0^{4\beta} \int_0^{4\beta} \dd{s_1} \dd{s_2} \chi_k(s_1) \left( \partial - \hat{\Sigma}(s_2,s_1) \right) \chi_k(s_2),
\end{align}
where $\hat{\Sigma}$, the self-energy of the measured fields, is a piecewise defined function
\begin{align}
    \hat{\Sigma}(s_1,s_2) &= \nonumber \\*
    & \begin{bmatrix}
         -\Sigma_{11}(s_1,s_2) & -\Sigma_{21} & \Sigma_{21} & -\Sigma_{11} &-\Sigma_{11}&  -\Sigma_{21} &  \Sigma_{21} & -\Sigma_{11} \\
         -\Sigma_{12}(\beta-s_1,s_2)&-\Sigma_{22}&\Sigma_{22}&-\Sigma_{12}&-\Sigma_{12}&-\Sigma_{22}&\Sigma_{22}&-\Sigma_{12} \\
         \Sigma_{12}(2\beta-s_1,s_2)&\Sigma_{22}&-\Sigma_{22}&\Sigma_{12}&\Sigma_{12}&\Sigma_{22}&-\Sigma_{22}&\Sigma_{12} \\
         -\Sigma_{11}(s_1 -\beta,s_2)&-\Sigma_{21}&\Sigma_{21}&-\Sigma_{11}&-\Sigma_{11}&-\Sigma_{21}&\Sigma_{21}&-\Sigma_{11} \\
         -\Sigma_{11}(s_1 -\beta,s_2)&-\Sigma_{21}&\Sigma_{21}&-\Sigma_{11}&-\Sigma_{11}&-\Sigma_{21}&\Sigma_{21}&-\Sigma_{11} \\
         -\Sigma_{12}(4\beta-s_1,s_2)&-\Sigma_{22}&\Sigma_{22}&-\Sigma_{12}&-\Sigma_{12}&-\Sigma_{22}&\Sigma_{22}&-\Sigma_{12} \\
         \Sigma_{12}(5\beta-s_1,s_2)&\Sigma_{22}&-\Sigma_{22}&\Sigma_{12}&\Sigma_{12}&\Sigma_{22}&-\Sigma_{22}&\Sigma_{12} \\
         -\Sigma_{11}(s_1 -2\beta,s_2)&-\Sigma_{21}&\Sigma_{21}&-\Sigma_{11}&-\Sigma_{11}&-\Sigma_{21}&\Sigma_{21}&-\Sigma_{11}
    \end{bmatrix}.
\end{align}
Note that that the integrals in (\ref{kin_chi}) and the arguments in $\hat{\Sigma}(s_1,s_2)$ now range from $0$ to $4\beta$. For $s_1 \in \left( (i-1)\frac{\beta}{2},i \frac{\beta}{2} \right)$ and $s_2 \in \left( (j-1)\frac{\beta}{2},j \frac{\beta}{2} \right)$, we read of the $i$'th row and $j$'th column, with $i,j = 1,...,8$, of the above matrix to get the value of the function $\hat{\Sigma}(s_1,s_2)$. The $s_{1/2}$ dependence is only written explicitly for the first column. The first argument of $\Sigma_{ab}$ in the $i$'th row and $j$'th column of $\hat{\Sigma}$ will be the same as the argument of the $i$'th entry in $\chi_k(s)$, whereas the second argument of  $\Sigma_{ab}$ will be the same as the argument of the $j$'th entry in $\chi_k(s)$ in (\ref{chi}).

From (\ref{kin_chi}), we observe that the kinetic term of the measured Dirac fermions is replaced by a Majorana like kinetic term for the $\chi$-field (on a doubled imaginary time contour), showing up with the right sign and prefactor.
After integrating out the fermions, we get the final form of the large-$N$ action in terms of the bilocal fields
\begin{align}\label{large_N_final}
    -\frac{I}{N} &= \frac{m}{2} \log\det\left(\partial - \hat{\Sigma}\right) + (1-m) \log \det(\partial - \Sigma_{33}) \nonumber\\
     &\hphantom{=}+\iint \dd{\tau_1} \dd{\tau_2}  -\frac{m}{4} \Sigma_{11}(\tau_1,\tau_2) G_{11}(\tau_2,\tau_1) -\frac{m}{4} \Sigma_{12}(\tau_1,\tau_2) G_{12}(\tau_2,\tau_1) \nonumber \\
    &\hphantom{=}  - \frac{m}{4} \Sigma_{21}(\tau_1,\tau_2) G_{21}(\tau_2,\tau_1) - \frac{m}{4} \Sigma_{22}(\tau_1,\tau_2) G_{22}(\tau_2,\tau_1) \\
    &\hphantom{=} - (1-m) \Sigma_{33}(\tau_1,\tau_2) G_{33}(\tau_2,\tau_1) \nonumber \\
    &\hphantom{=}  + \frac{J^2}{q} \left(-\frac{m}{4} G_{11}(\tau_1,\tau_2) - \frac{m}{4} G_{12} + \frac{m}{4} G_{21} + \frac{m}{4} G_{22}  + (1-m)G_{33} \right)^{\frac{q}{2}} \nonumber \\
    & \hphantom{+\frac{J^2}{q}\Big(}\!\!\!\times \left(-\frac{m}{4} G_{11}(\tau_2,\tau_1) - \frac{m}{4} G_{12} + \frac{m}{4} G_{21} + \frac{m}{4} G_{22}  + (1-m)G_{33} \right)^{\frac{q}{2}} \nonumber, 
\end{align}
where $m\equiv M/N$. It is useful to introduce a two-point-function for all the measured fermions $\hat{G}(s_1,s_2)$
\begin{align}
   \int_0^{2\beta} \int_0^{2\beta} \dd{\tau_1} \dd{\tau_2} \frac{m}{4} &\left[ - \Sigma_{11}(\tau_1,\tau_2) G_{11}(\tau_2,\tau_1) - \Sigma_{12}(\tau_1,\tau_2) G_{12}(\tau_2,\tau_1) \nonumber \right. \\
    &\left. -  \Sigma_{21}(\tau_1,\tau_2) G_{21}(\tau_2,\tau_1) -  \Sigma_{22}(\tau_1,\tau_2) G_{22}(\tau_2,\tau_1) \right] \\
    = \int_0^{4\beta} \int_0^{4\beta} \dd{s_1} \dd{s_2} & \:
    \frac{m}{2} \left( -\hat{\Sigma}(s_1,s_2) \hat{G}(s_2,s_1) \right), \nonumber
\end{align}
where
\begin{align}\label{G_hat}
    \hat{G}(&s_1,s_2) = \frac{1}{M} \sum_{k=1}^{M} \chi_k(s_1) \chi_k(s_2) \nonumber \\
    =& - \frac{1}{2} \left[\begin{array}{rrrrrrrr}
         G_{11}(s_1,s_2) & G_{12} &- G_{12} & G_{11} &G_{11}&  G_{12} & - G_{12} & G_{11} \\
         G_{21}(\beta-s_1,s_2)&G_{22}&-G_{22}&G_{21}&G_{21}&G_{22}&-G_{22}&G_{21} \\
         -G_{21}(2\beta-s_1,s_2)&-G_{22}&G_{22}&-G_{21}&-G_{21}&-G_{22}&G_{22}&-G_{21} \\
         G_{11}(s_1 -\beta,s_2)&G_{12}&-G_{12}&G_{11}&G_{11}&G_{12}&-G_{12}&G_{11} \\
         G_{11}(s_1 -\beta,s_2)&G_{12}&-G_{12}&G_{11}&G_{11}&G_{12}&-G_{12}&G_{11} \\
         G_{21}(4\beta-s_1,s_2)&G_{22}&-G_{22}&G_{21}&G_{21}&G_{22}&-G_{22}&G_{21} \\
         -G_{21}(5\beta-s_1,s_2)&-G_{22}&G_{22}&-G_{21}&-G_{21}&-G_{22}&G_{22}&-G_{21} \\
         G_{11}(s_1 -2\beta,s_2)&G_{12}&-G_{12}&G_{11}&G_{11}&G_{12}&-G_{12}&G_{11} 
    \end{array}\right].
\end{align}
The Schwinger-Dyson equations are derived by setting variations of the action with respect to $\hat{\Sigma}$, $G_{ab}$ and $G_{33}$ to zero
\begin{equation}
    \begin{aligned}
    \hat{G} =& (\hat{G}_{\text{free}}^{-1} - \hat{\Sigma})^{-1}, \\
    G_{33} =& (G_{33,\text{free}}^{-1} - \Sigma_{33})^{-1}, \\
    \Sigma_{11}(\tau_1,\tau_2) =& \Sigma_{12} = - \Sigma_{21} = - \Sigma_{22} = -\Sigma_{33} 
    \\ =& -J^2\left(-\frac{m}{4} G_{11}(\tau_1,\tau_2) + \frac{m}{4} G_{22} - \frac{m}{4} G_{12} + \frac{m}{4} G_{21} + (1-m)G_{33}\right)^{\frac{q}{2}} \\
    &\hphantom{-\;\,}\times\left(-\frac{m}{4} G_{11}(\tau_2,\tau_1) + \frac{m}{4} G_{22} - \frac{m}{4} G_{12} + \frac{m}{4} G_{21} + (1-m)G_{33}\right)^{\frac{q}{2}-1}.
\end{aligned}\label{schwinger_q}
\end{equation}
Here, $i,j \in \left\{1,2\right\} \lor i=j = 3$. Notice that the free propagator is the inverse of the kinetic operator in the non-interacting theory. The explicit form of the free propagators $\hat{G}_{\text{free}}$ and $G_{33,\text{free}}$ depends on the boundary conditions, see (\ref{bc_meas_num}) -- (\ref{bc_unmeas_den}). This is important when solving the Schwinger-Dyson equations numerically and will be subject of the next subsection.

\subsection{R\'{e}nyi-2 entropy}\label{section_renyi_2}
We plan to solve the Schwinger-Dyson equations (\ref{schwinger_q}) numerically by an iterative approach and plug the solutions back into the action, to calculate the entanglement entropy $S_R/N$ via (\ref{entropy_saddle_point}).

We start by solving the equations of motion for the numerator with boundary conditions (\ref{bc_meas_num}) and (\ref{bc_unmeas_num}). These boundary conditions ensure continuity of $\chi_k(s)$ in (\ref{chi}) in the region $s \in (\beta,3\beta)$, with anti-periodicity $\chi_k(\beta_+) =- \chi_k(3\beta_-)$, as well as the region $s \in (0,\beta) \cup (3\beta,4\beta)$ with $\chi_k(0) = -\chi_k(4\beta)$. The corresponding propagator can therefore be interpreted as two fermions propagating freely on two separate contours. The unmeasured fermions, because of the boundary conditions arising from the twist operator $c_i(0) = -c_i(2\beta)$, have the usual free propagator on the full contour of the replicated geometry. The free propagators for the numerator boundary conditions therefore take the following form (analogously to \cite{Antonini_2023})
\begin{align}
    &\hat{G}_{\text{free}}(s_1,s_2) =
    \begin{cases}
    \frac{1}{2} \text{sgn}(s_1-s_2), & s_1,s_2 \in (\beta,3\beta) \lor s_1,s_2 \in (0,\beta) \cup (3\beta,4\beta), \\
    0, & \text{else,}
    \end{cases} \label{G_free_num_hat} \\
    &G_{33,\text{free}}(\tau_1,\tau_2) = \frac{1}{2} \text{sgn}(\tau_1-\tau_2), \;\;\;\; \tau_1,\tau_2 \in (0,2\beta). \label{G_free_num_33}
\end{align}
Now, everything is set to solve the Schwinger-Dyson equations iteratively, for details see for example \cite{Maldacena_2016,Zhang_2020}.

By plugging in the solutions to the Schwinger-Dyson equations (\ref{schwinger_q}) in the large-$N$ action (\ref{large_N_final_q}), one arrives at the on-shell action
\begin{align}\label{on_shell_num}
    -\frac{I^\ast_{num}}{N} = &\frac{m}{2} \left( \Tr \log  \left[ \left(\hat{G}_{\text{free}}^{-1} - \hat{\Sigma}\right) \hat{G}_{\text{free}} \right] + 2\log 2 \right) \nonumber \\
    +& (1-m) \left( \Tr \log \left[ \left (G_{33,\text{free}}^{-1} - \Sigma_{33} \right) G_{33,\text{free}} \right] + \log 2 \right)\\
    + &\iint \dd{\tau_1} \dd{\tau_2} J^2 \left(\frac{1}{q} - 1\right) \abs{-\frac{m}{4} G_{11} - \frac{m}{4} G_{12} + \frac{m}{4} G_{21} + \frac{m}{4} G_{22}  + (1-m)G_{33} }^q,  \nonumber
\end{align}
where we used the identity $\log \det\mathcal{O} = \Tr \log\mathcal{O}$ and the normalisation for the free propagators
\begin{align}
    \Tr \log \hat{G}_{\text{free}}^{-1} = 2\log2, &&  \Tr \log G_{33,\text{free}}^{-1} = \log 2, \hfill
\end{align}
as in \cite{Antonini_2023}. Notice that the different normalisations appear because $G_{33,\text{free}}$ represents the propagator of a free fermion on a contour $s \in (0, 2\beta)$, while $\hat{G}_{\text{free}}$ represents the propagator of two free fermions on two separate contours $s \in (\beta, 3\beta)$ and $s \in (0, \beta) \cup (3\beta, 4\beta)$.

The same can be done for the denominator boundary conditions (\ref{bc_meas_den}) and (\ref{bc_unmeas_den}). Now $\chi_k(s)$ is continuous in $s \in (0,2\beta)$, with anti-periodicity $\chi_k(0) = - \chi_k(2\beta_-)$, as well as in $s \in (2\beta , 4\beta)$ with $\chi_k(2\beta_+) = - \chi_k(4\beta)$. This leads to the free propagators
\begin{align}
    \hat{G}_{\text{free}} =&
    \begin{cases}
    \frac{1}{2} \text{sgn}(s_1-s_2), & s_1,s_2 \in (0,2\beta) \lor s_1,s_2 \in (2\beta,4\beta), \\
    0, & \text{else,}
    \end{cases} \label{G_free_den_hat} \\
    G_{33,\text{free}} = &
    \begin{cases}
    \frac{1}{2} \text{sgn}(s_1-s_2), & s_1,s_2 \in (0,\beta) \lor s_1,s_2 \in (\beta,2\beta), \\
    0, & \text{else.}
    \end{cases} \label{G_free_den_33}
\end{align}
Then, the solutions obtained by the iterative procedure can be plugged into the action
\begin{align}\label{on_shell_den}
    -\frac{I^\ast_{den}}{N} =& \frac{m}{2} \left( \Tr \log  \left[ \left(\hat{G}_{\text{free}} - \hat{\Sigma}\right) \hat{G}_{\text{free}} \right] + 2\log 2 \right) \nonumber\\
    + &(1-m) \left( \Tr \log \left[ \left (G_{33,\text{free}} - \Sigma_{33} \right) G_{33,\text{free}} \right] + 2\log 2 \right) \\
    + &\iint \dd{\tau_1} \dd{\tau_2} J^2 \left( \frac{1}{q} - 1 \right) \abs{-\frac{m}{4} G_{11} - \frac{m}{4} G_{12} + \frac{m}{4} G_{21} + \frac{m}{4} G_{22}  + (1-m)G_{33} }^q.  \nonumber
\end{align}
\\
Note that the normalisation of both free propagators is given by $\Tr \log G^{-1} = 2 \log 2$, corresponding to two fermions on separate contours.

Finally, the R\'{e}nyi-2 entropy can be computed as a function of the measured fermions by simply plugging in the solution of the Schwinger-Dyson equations in $I^\ast_{num}$ and $I^\ast_{den}$ and then computing \eqref{entropy_saddle_point}.

\begin{figure}
 \centering
\begin{subfigure}{0.75\textwidth}
    
    \includegraphics[width=\textwidth]{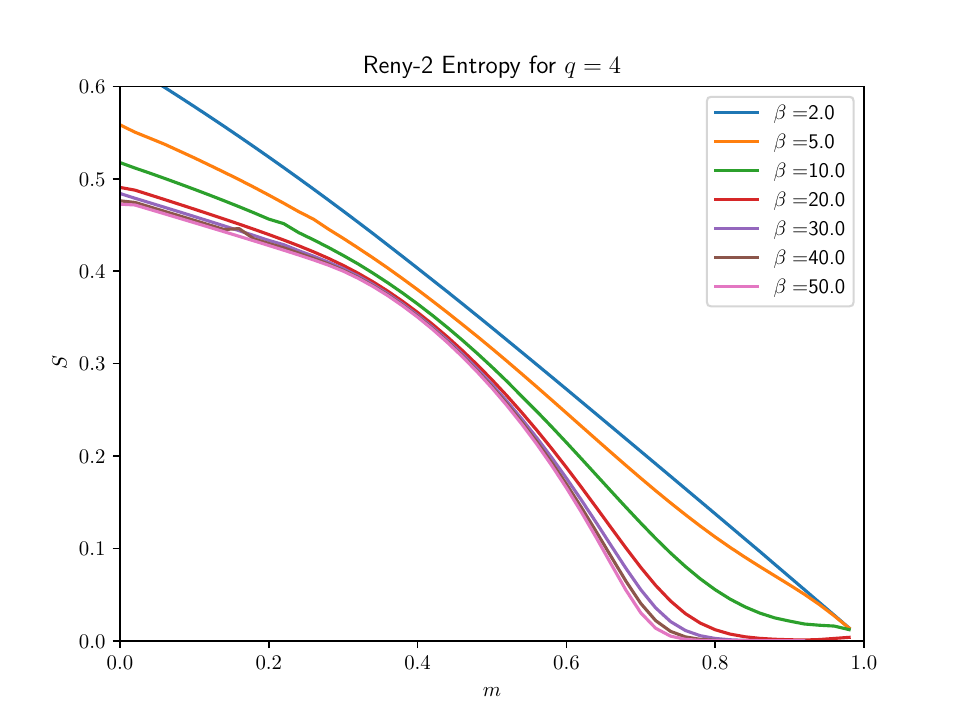}
    \subcaption{R\'{e}nyi-2 entropy plotted against $m=M/N$.}
    \label{fig:reny_2_vs_m}
\end{subfigure}
\begin{subfigure}{0.75\textwidth}
    \includegraphics[width=\textwidth]{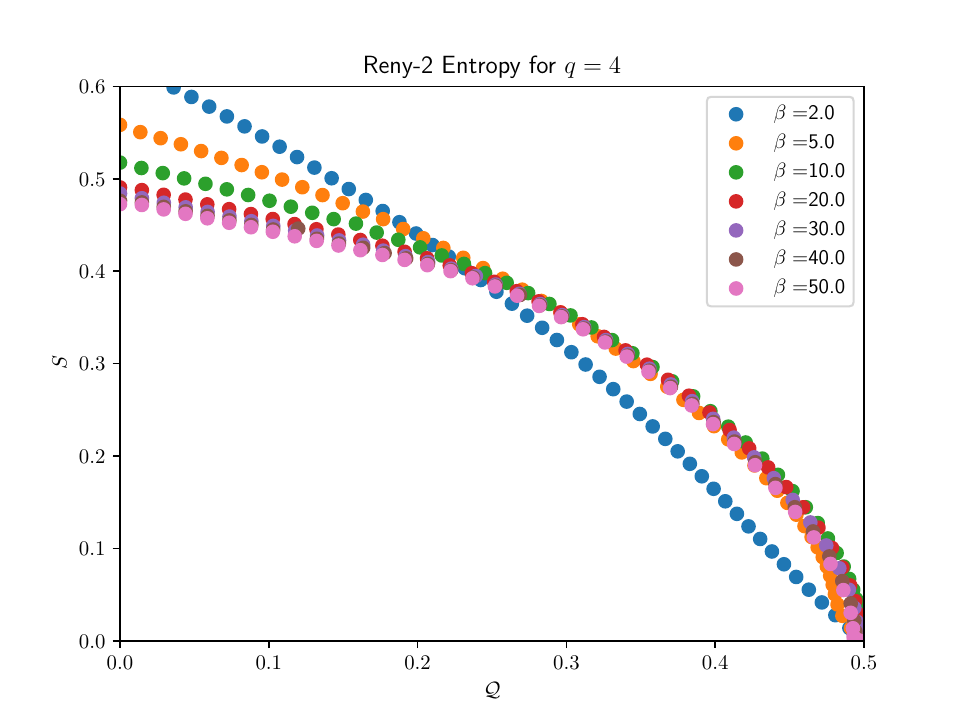}
    \subcaption{R\'{e}nyi-2 entropy plotted against $\mathcal{Q}$.}
    \label{fig:reny_2_vs_Q}
\end{subfigure}
\label{fig:reny_2_numerical}
\caption{Numerical results for the entropy of the setup described in \cref{section_syk_measurement}. $m=M/N$ is the ratio of measured fermions to their total number. $\mathcal{Q}$ is the relative charge of the system.}
\end{figure}

The entanglement entropy as a function of the number of measured fermions $m$ is plotted for different $\beta$ in \cref{fig:reny_2_vs_m}. Looking at the curves, we can see that for small $\beta$ the curves look almost linear. This is to be expected as in this limit the system will behave just like an ensemble of free fermions. However, as we go to larger $\beta$, the curves' shapes get more nuanced and we can differentiate between three major sections: In the first section the curves are more or less linear. Around $m=0.4$, we then encounter a section of rapid change. The entropy quickly drops towards zero and finally in the last section from around $m=0.8$ almost flatlines at nearly vanishing entropy.

To understand this behaviour, it is important to understand the phase structure of the cSYK first. This will mainly be based on the results of \cite{cao2021thermodynamic}, although we also reproduce some of their results in \cref{app: Phase structure}. 
The cSYK at $q=4$ has two stable phases, which we shall refer to as liquid and gaseous. The gaseous phase occurs at small charge density and high temperature. It is characterised by low correlation between the single site charges. Increasing the charge density by raising the chemical potential, or as we will see by charge measurement, may lead to a phase transition, if the temperature is below the critical temperature. In the liquid phase, the single site charges align which leads to a jump in the charge density and in consequence a drop in entropy. A phase transition of this kind may explain the behaviour we see in \cref{fig:reny_2_vs_m}. To investigate this further, we will also take a look at the charge curve.

\subsection{Comparison to the thermal partition 
function }

In preparation for the bulk calculations, here we want to give the partition function after measurement and compare it to its thermal version in the canonical ensemble.

The Hilbert space for the cSYK model admits two different bases that we will make use of and switch between to make our argument. The first basis, in terms of energy and charge, has already been used above to define the TFD state (see \cref{TFD_cSYK}). Since $\left[ H, Q\right] = 0$, charge and energy are good quantum numbers. Notice that a charge flip $Q\rightarrow -Q$ does not change the energy of a state, due to the particle-hole symmetry of the Hamiltonian (when $\mu =0$). Thus, states with finite charge are (at least) twofold degenerate with respect to $H$ and any charge subsector admits the exact same energy spectrum as its counterpart of the opposite charge \cite{Antonini_2021}.

The second basis we will employ is comprised of eigenstates of the charge at site $k$ operators $\left\{ Q_k\right\}$ introduced in \cref{eq:Charge_at_site_k}. We will classify these states in terms of their total charge $Q$ and the respective permutation of the single fermion charge, i.e.\ the basis for each charge subsector with charge eigenvalue Q is given by the set
\begin{equation}
    B_Q \equiv \left\{ \ket{P_i( \underbrace{\uparrow \dots \uparrow }_{N\left(\frac{1}{2}+Q\right)} \underbrace{\downarrow \dots \downarrow}_{N\left(\frac{1}{2}-Q\right)})} \right\}
\end{equation}
where each of the $P_i$s is one specific permutation of its arguments and $i\in[1,D_Q]$, where $D_Q$ is the dimension of the respective charge subsector. We shall use the more concise notation $\ket{P_i( \uparrow \dots \uparrow \downarrow \dots \downarrow)} \equiv \ket{P_i(Q)}$.

The two bases are related to each other via (see \cite{Antonini_2021})
\begin{equation}
  \ket{P_i(Q)}  = \sum^{D_Q}_j c_j^i \ket{E_j,Q} \label{eq:basis_change}
\end{equation} 
which implies
\begin{equation}
   1 \stackrel{!}{=} \sum^{D_Q}_j  \left| c_j^i\right|^2,
\end{equation}
for the states $\ket{P_i(Q)}$ to be normalized properly. Other than that we cannot say anything about the $c_j^i$ in general. However, we only care about the averaged case. After averaging over the couplings $\Ji$, none of the permutations are special and we can therefore conclude that 
\begin{equation}
\overline{\left| c_j^i\right|^2} = \frac{1}{D_Q}. \label{eq:averaged_phase}
\end{equation}
\\
We can now finally turn to the partition function. After measurement, we find
\begin{equation}
    \begin{aligned}
    \overline{Z}_m &\propto \Tr \left( e^{-\beta \overline{H}} \ketbra{L(M)}{L(M)}\right) \\
    &= \sum_{Q=-1/2}^{1/2}\sum^{D_Q}_i \bra{E_i, Q}  e^{-\beta \overline{H}} \ketbra{L(M)}{L(M)} E_i, Q\rangle \\
    &= \sum_{Q=-1/2}^{1/2}\sum^{D_Q}_i \bra{P_i(Q)}  e^{-\beta \overline{H}} \ketbra{L(M)}{L(M)} P_i(Q)\rangle\\
    &= \sum_{Q=\frac{2m-1}{2}}^{1/2}\sum_{P_i \in B_Q^M} \bra{P_i(Q)}  e^{-\beta \overline{H}} \ket{P_i(Q)} \\
    &= \sum_{Q=\frac{2m-1}{2}}^{1/2}\sum_{P_i \in B_Q^M} \frac{1}{D_Q}\sum^{D_Q}_{j} \bra{E_j, Q}  e^{-\beta E_j} \ket{E_j, Q},
\end{aligned}
\end{equation}
where in the second to last line we introduced the restricted basis of the subspace that the charge subsector is projected onto,
\begin{equation}
    B_Q^M = \left\{ P_i(Q) \in B_Q : P_i(Q) = \ket{\underbrace{\uparrow\dots \uparrow }_M \dots } \right\},
\end{equation}
and in the last line we used \cref{eq:basis_change,eq:averaged_phase}. With that, the sum over permutations decouples and defining $D_{Q}^m \equiv \dim(B_Q^M)$, we can now write
\begin{align}
    Z_m &\propto \sum_{Q=\frac{2m-1}{2}}^{1/2} \frac{D_{Q}^m}{D_Q}\sum^{D_Q}_{j}  e^{-\beta E_j} .\label{eq:Z_m}
\end{align}
The coefficient $D_{Q}^m$ can be obtained through simple combinatorics and reads
\begin{align}
    D_{Q}^m = \binom{N-M}{N(\frac{1}{2}+\mathcal{Q})-M} = \frac{(N(1-m))!}{(N(\mathcal{Q}+\frac{1}{2}-m))!(N(\frac{1}{2}-\mathcal{Q}))!}
\end{align}
for $m\ge Q+1/2$ and $D_{Q}^m = 0$ otherwise. For $m=0$, we recover $D_Q^0=D_Q$ as expected and
for small $m$, we can thus expand
\begin{equation}
   \frac{D_{Q}^m}{D_Q} =  1 + m\log\left(\frac{1}{2} + \mathcal{Q}\right) + \mathcal{O}(m^2),
\end{equation}
It is easy to prove that for $\beta \rightarrow 0$, where all energies are equally probable, the system gets projected onto the $Q=m/2$ charge subsector exactly. However, in the finite $\beta$ case, the resulting charge after measurement depends on the weights $\exp{-\beta E_j}$, which are at this point unknown to us. Nevertheless, we know that the ground state lies within the $\mathcal{Q} = 0$ subsector and the subsequent sectors with larger $|\mathcal{Q}|$ also have successively larger energies \cite{Antonini_2021}. For small $m$ and large $\beta$ we can therefore focus on the small $Q$ subsectors and approximate
\begin{equation}
    \log\left(\frac{1}{2}+\mathcal{Q}\right) \approx 2\mathcal{Q}-\log 2 + \mathcal{O}(Q^2).
\end{equation}
Equipped with this approximation, we can give an estimate for our modified partition function,
\begin{equation}
    Z_m = \sum_{Q=\frac{2m-1}{2}}^{1/2} \sum^{D_Q}_{j} e^{-\beta(E_j -\frac{2mQ}{\beta}) -m\log 2},
\end{equation}
In turn, this immediately yields an expression for the entropy after measurement for small $m$
\begin{equation}
    S_m(Q,\beta) = \braket{E}_m + \mu_\text{eff.}(m,\beta)\braket{Q} - \ln(Z) + m S_\text{bdry.},
\end{equation}
where we introduced the effective chemical potential $\mu_\text{eff.}$, a function of $m$, and the constant boundary entropy $S_\text{bdry.}$. But, this is just the grand canonical entropy plus some boundary term linear in $m$, which confirms the BCFT result \cite{Takayanagi_2011,cavalcanti2019studiesboundaryentropyadsbcft}.

\subsection{Charge after measurement} \label{sec:charge_after_measurement}
\begin{figure}[h!]
    \includegraphics[width=\textwidth]{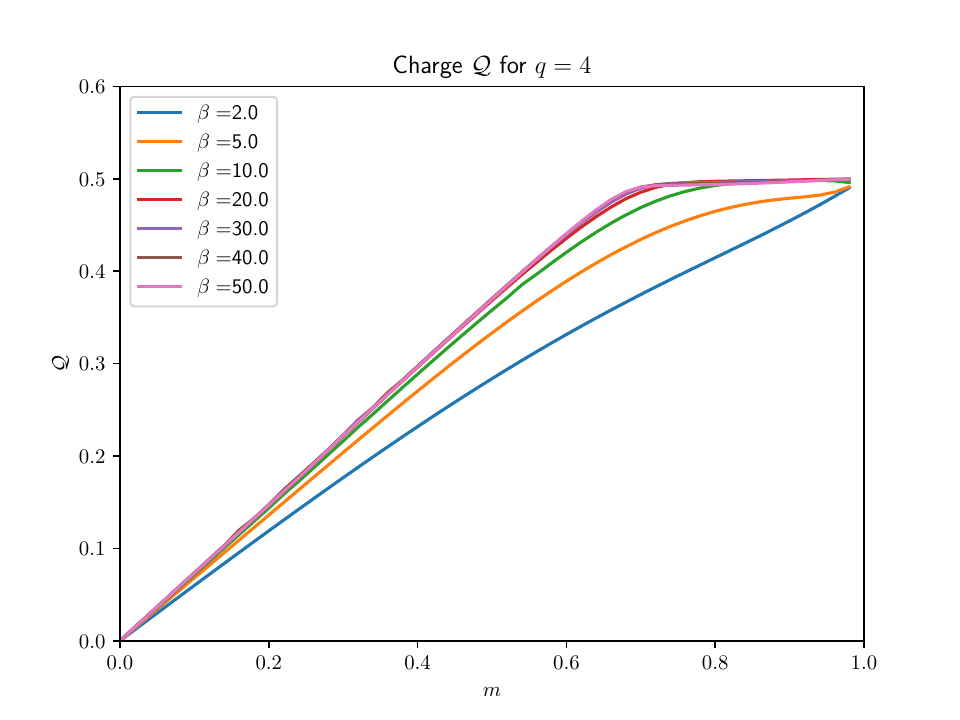}
    \caption{Numerical result for the relative charge $\mathcal{Q}$ in the setup described in \cref{section_syk_measurement} plotted against $m$.}
    \label{fig:Qvsm}
\end{figure}

As we have seen in the previous section, the process of measuring gives our system a finite net charge $\braket{\mathcal{Q}}_m$, by blocking out certain modes from the spectrum. In the bulk, this charge will be dual to the charge of the black hole \cite{Davison_2017,Gaikwad_2020,cao2021thermodynamic}. In order to calculate the bulk entropy, we will thus need an expression for $\braket{\mathcal{Q}}_m$ in terms of cSYK propagators. As before, the subscript $m$ implies that boundary conditions due to the measurement of the first $M=mN$ fermions must be imposed, i.e. $\braket{\dots}_m = Z_m^{-1}\braket{\dots \ketbra{L_M}{L_M}}_\beta$. We have
\begin{equation}
\begin{aligned}
    \braket{\mathcal{Q}}_m &= \frac{1}{N} \sum_{k=1}^N\braket{c_k^\dagger c_k}_m - \frac{1}{2} \\
    &= \frac{1}{N}\left( \sum_{k=1}^M\braket{c_k^\dagger c_k}_m + \sum_{k=M+1}^N\braket{c_k^\dagger c_k}_m\right) - \frac{1}{2} \\
    &=  \frac{1}{N}\left( M + \sum_{k=M+1}^N\braket{c_k^\dagger c_k}_m\right) - \frac{1}{2}, \label{eq:mean_charge_1}
\end{aligned}
\end{equation}
where in the last line we used that $c_k^\dagger c_k$ for the first $M$ fermions has eigenvalue $1$. Before we proceed, we note that $c_k^\dagger c_k$ commutes with the Hamiltonian and thus $c_k^\dagger c_k = c_k^\dagger (\tau_0) c_k(\tau_0)$. Since the propagators $G_{ij}(\tau_1, \tau_2)$ are ill-defined at $\tau_1=\tau_2$, it will be useful to introduce an infinitesimal offset $\varepsilon$ to the time variable of the $c_k^\dagger$ operator:
\begin{equation}
    \begin{aligned}
 T\left[c_k(\tau_0)c_k^\dagger(\tau_0-\varepsilon)\right] &= c_k(\tau_0)c_k^\dagger(\tau_0-\varepsilon) \\
 &= c_k(\tau_0)U^{\dagger}(-\varepsilon)c_k^\dagger (\tau_0)U(-\varepsilon) \label{eq:small_time_shift}\\
 &= c_k(\tau_0)(1-i\varepsilon H)c_k^\dagger(\tau_0)(1+i\varepsilon H) \\
 &= c_k(\tau_0) c_k^\dagger(\tau_0) + i  \varepsilon c_k\left[ c_k^\dagger, H\right] \\
 &= 1-c_k^\dagger c_k + i  \varepsilon c_k \left[ c_k^\dagger, H\right]
\end{aligned}
\end{equation}
and similarly
\begin{equation}
    \begin{aligned}T\left[c_k(\tau_0)c_k^\dagger(\tau_0+\varepsilon)\right] &= -c_k^\dagger(\tau_0+\varepsilon)c_k(\tau_0) \\
 &= -c_k^\dagger c_k + i \varepsilon \left[ c_k^\dagger, H\right] c_k  \\
 &= -c_k^\dagger c_k -  i \varepsilon c_k \left[ c_k^\dagger, H\right] ,
\end{aligned}
\end{equation}
where in \cref{eq:small_time_shift} we introduced the time evolution operator $U$. Thus,
\begin{align}
   c_k^\dagger c_k = \frac{1}{2} \left( 1 -T\left[c_k(\tau_0)c_k^\dagger(\tau_0-\varepsilon)\right] - T\left[c_k(\tau_0)c_k^\dagger(\tau_0+\varepsilon)\right]\right)
\end{align}
and with that and using the expression for $G_{33}$ in terms of $c_k$ and $c_k^\dagger$, we can finally write
\begin{align}\label{eq:charge_after_meas}
    \braket{\mathcal{Q}}_m = \frac{m}{2} + \frac{1-m}{2}\left( G_{33}(\tau_0, \tau_0-\varepsilon) +  G_{33}(\tau_0, \tau_0+\varepsilon) \right).
\end{align}
Here, $\tau_0$ is just any point in the range $[0,\beta]$. We can now use the above expression to calculate $\braket{\mathcal{Q}}$ from our numerical results. The corresponding curve as a function of $m$ can be found in \cref{fig:Qvsm}. \Cref{fig:reny_2_vs_Q} again shows the R\'{e}nyi-2 entropy for our setup, but this time plotted against $\mathcal{Q}$. As we can see in \cref{fig:Qvsm}, the charge behaves as we had expected. For large $\beta$ the curve suddenly flattens out around $m\approx0.8$, affirming our believe that a phase transition occurs around that point. The argument here is the following: After a certain threshold of measurements is completed, the single site charges suddenly become correlated and align. Further measurements therefore do not have any effect leading to the flat curve. Comparing \cref{fig:reny_2_vs_Q} to \cref{fig:gc_results_v_charge}, confirms this viewpoint. Qualitatively, the entropy curves with measurement are almost identical to the ones with a chemical potential where we know the phase transition takes place (the difference in $S(Q=0)$ are very likely due to the fact that we are comparing von Neumann and Rényi-2 entropy, see \cref{sec:neumannvrenyi}). We are now ready to proceed to the holographic dual of our setup.

\section{Gravity Dual}
\label{sec:bulk}
In this section, we construct the holographic gravity dual of the charge measurement on the cSYK TFD. We make use of the quantum extremal surface (QES) \cite{Engelhardt_2015} procedure to show that the entropy curve of a charged black hole reproduces the behaviour observed in the boundary theory. This enables us to choose the correct entropy associated to the QES as well as the location of the entanglement wedge based on the number of boundary fermions being measured.

\subsection{Entropy in the bulk}\label{section_bulk}
Upon acting with the measurement operator \eqref{eq:Charge_at_site_k} on the cSYK (asymptotic boundary theory), boundary conditions \eqref{bc_meas_num} --\eqref{bc_unmeas_den} are imposed on all the Dirac fermions at the instance when the measurement is performed. These boundary conditions allow us to treat the cSYK model as a ``BCFT''. Following \cite{Takayanagi_2011}, an end of the world (ETW) brane anchored at the asymptotic boundary at the same instance when the measurement operator is implemented is produced within the bulk. The ETW brane extends into the bulk and has Neumann boundary conditions imposed on it. Such an ETW brane is only visible to the measured cSYK fermions and cuts off part of the bulk accessible to these measured fermions. A consequence of the ETW brane and its boundary conditions is that the original bulk CFT gets split such that there are $N-M$ degrees of freedom in the bulk CFT dual to the unmeasured Dirac fermions while $M$ degrees of freedom in the bulk BCFT dual to the measured Dirac boundary fermions. Following \cite{Gaikwad_2020,Antonini_2021}, we dimensionally reduce the bulk theory and impose the appropriate boundary conditions on the asymptotic boundary and the ETW brane surface. Particularly, boundary conditions are necessary to avoid charge fluctuations. The resulting two dimensional dilaton theory resembles JT gravity coupled to a gauge field and is well defined.

Similar to \cite{Antonini_2023}, we propose the following dual setup of the boundary theory discussed in the previous section: a two dimensional gauge field plus JT gravity (see \cref{theory_dual_cSYK}) coupled to a CFT of $N-M$ free fermions (dual to the unmeasured fermions) and a BCFT of $M$ free fermions (dual to the measured fermions). 

The equations of motion of the dimensionally reduced dual bulk theory in the presence of an energy-momentum tensor $T_{\mu\nu}$, associated to the matter fields (gauge field + CFT/BCFT) are (setting the radius of curvature $l_{\text{AdS}}=1$)
\begin{align}
    R  &= -2,\\
    T_{\mu\nu} &= \frac{1}{8\pi G_2} \left( \nabla_\mu \nabla_\nu \phi - h_{\mu\nu} \nabla^2 \phi + \frac{h_{\mu\nu}}{l^2} \phi \right). 
\end{align}
At inverse temperature $\beta$ (setting $T_{\mu\nu}=0$), 
the metric and dilaton profile are given by
\cite{Antonini_2023}
\begin{align}
    \dd s^2 = \frac{4\pi^2}{\beta^2} \frac{\dd \sigma^2 + \dd \tau^2}{\sinh^2 \frac{2\pi}{\beta}\sigma}, \label{eq: metric sol}\\
    \phi(\sigma) = \phi_r \frac{2\pi}{\beta} \frac{1}{\tanh \frac{2\pi}{\beta}\sigma}.\label{eq: dilaton sol}
\end{align}
The $(\tau,\sigma)$ coordinates only cover one exterior region of the double sided black hole. In these coordinates the horizon lies at $\sigma = \infty$ and the asymptotic boundary at $\sigma = \epsilon$. Therefore, the dilation profile at the horizon is
\begin{equation}\label{eq: dilaton at horizon}
    \phi = \phi_r\frac{2\pi}{\beta}.
\end{equation}

The entropy of some subset $R$ of the boundary system can be calculated with the quantum extremal surface (QES) \cite{Engelhardt_2015,penington2020replica} formula
\begin{align}\label{eq: QES formula}
    S(R) = \text{min}\left\{ \text{ext}_\sigma \; S_{\text{Gen}}(\sigma) \right\},
\end{align}
where $S_{\text{Gen}}(\sigma)$ is the generalised entropy in a region bounded by $\sigma$ and the asymptotic boundary. $S_{\text{Gen}}(\sigma)$ has an area contribution plus a bulk matter entropy contribution. The area contribution is replaced by the value of the dilaton at $\sigma$ in JT-gravity \cite{Almheiri_2020}. From \cref{theory_dual_cSYK}, one gets an additional term from the on-shell action of the gauge field contribution. 
The entropy in a region bounded by $\sigma$ is given by
\begin{align}\label{S_gen}
    S_{\text{Gen}}(\sigma) &= S_0 + S_{\text{Dyna}}(\sigma) + S_{\text{Gauge}}+ (N-M) S_{\text{CFT}} + M S_{\text{BCFT}}.
\end{align}
The first term in equation (\ref{S_gen}) originates from the topological Einstein-Hilbert term (which also corresponds to the ground state entropy) proportional to the background value of the dilaton $\phi_0$, the second term is the first correction linear in temperature and is the only $\sigma$ dependent term (see \eqref{eq: dilaton sol}). The third term is the on-shell gauge field contribution. The last term denotes the entropy of the CFT/BCFT matter fields, modelling the dual unmeasured/measured fermionic fields on the boundary. Since the cSYK model describes Dirac-fermions, we model the CFT as free (massless) Dirac fermions and its associated entropy is given by \cite{Antonini_2023,Callebaut_2023}
\begin{align}\label{entropy_cft_bulk}
    S_{\text{CFT}} = \frac{c}{6} \log 2.
\end{align}
Here, $c$ denotes the central charge of a free Dirac fermion, hence $c=1$. The entropy of the BCFT is that of the CFT plus a term $\log g$ called the boundary entropy \cite{Antonini_2023,PhysRevLett.67.161}
\begin{align}\label{BCFT_entropy}
    S_{\text{BCFT}} = \frac{c}{6} \log 2 + \log g.
\end{align}

The boundary entropy $\log g$ is fixed analogous to \cite{Antonini_2023}: If we imagine the bulk matter to be dual to $N$ decoupled EPR pairs, then a measurement of one partner destroys the entanglement, which in turn leads to a vanishing entanglement entropy. Therefore, the boundary entropy should exactly cancel the ordinary CFT entropy contribution. From equation (\ref{BCFT_entropy}) we can therefore deduce $\log g = - \frac{\log 2}{6}$.

We can rewrite the generalised entropy slightly, by defining the ground state entropy of the system $\Tilde{S_0} = S_0 + N S_{\text{CFT}}$ such that,
\begin{align}\label{S_gen2}
    S_{\text{Gen}} &= \Tilde{S_0} + S_{\text{Dyna}} +S_{\text{Gauge}} + M \log g.
\end{align}

To establish a connection with the cSYK calculation discussed in the preceding section, the JT coefficients are determined by the identifications in (\ref{id_phi_r})--(\ref{ide_G_2}). Matching the ground state entropies of both theories yields
\begin{equation}
    \Tilde{S_0} = S_{0,\text{cSYK}}.
\end{equation}
Performing a large $q$ expansion, analogous to what was done in \cite{Davison_2017}, we find
\begin{align}
    S_{0,\text{cSYK}} = &\frac{2 \pi ^4 \left(-112 Q^4+24 Q^2+1\right)}{15 q^5}+\frac{\pi ^4 \left(48 Q^4-8 Q^2-1\right)}{12 q^4}+\frac{2 \pi ^2 \left(1-4 Q^2\right)}{3 q^3}\nonumber\\
    &+\frac{\pi ^2 \left(4 Q^2-1\right)}{2 q^2}+\frac{1}{2} \log \left(\frac{4}{1-4 Q^2}\right)+Q \log \left(\frac{2}{2 Q+1}-1\right).
\end{align}
The precise expressions for $S_{\text{Dyna}}$ and $S_{\text{Gauge}}$ can be obtained from the on-shell value of the charged JT gravity action via the expression
\begin{equation}\label{eq: S modified}
    S_{\text{Dyna}}+S_{\text{Gauge}} = \left[\beta\left(\partial_{\beta}-\left(\frac{\del\mu_\text{eff.}}{\del\beta}\right)\left(\frac{\del\mu_\text{eff.}}{\del Q}\right)^{-1}\right)-1\right]I^*. 
\end{equation}
This is obtained from the standard von Neumann entropy expression by performing the Legendre transformation from $\mu$ to $Q$. To calculate $S$, according to \cref{id_phi_r,id_K,iden_B,ide_G_2} we need the cSYK parameters $K$, $\gamma$ and $\mathcal{E}$. We compute $K$ and $\gamma$ numerically in \cref{sec:K_and_gamma}. A large $q$ expression for $\mathcal{E}$ can be obtained from \cref{eq: S modified} via \cref{curlyE},
\begin{align}
    \mathcal{E} = \frac{\frac{16 \pi ^3 Q}{5}-\frac{448 \pi ^3 Q^3}{15}}{q^5}+\frac{8 \pi ^3 Q^3-\frac{2 \pi ^3 Q}{3}}{q^4}-\frac{8 \pi  Q}{3 q^3}+\frac{2 \pi  Q}{q^2}+\frac{\log \left(\frac{2}{2 Q+1}-1\right)}{2 \pi }.
\end{align}
Note that before any measurement is performed, the complete state of the doubled system corresponds to the TFD state (\ref{TFD_cSYK}), which has (mean) charge zero, i.e.\ $\expval{\mathcal{Q}} = 0$. As $M$ fermions are measured, the mean charge increases to \eqref{eq:charge_after_meas}. This is important, as $\Tilde{S_{0}},\;S_{\text{Dyna}} \;\text{and} \;S_{\text{Gauge}}$ showing up in $S_{\text{gen}}$ are functions of the charge and therefore also functions of the number of measured fermions $m$.

The scenario when the QES is very close to the asymptotic boundary where the measurement operator acts needs an alternate treatment as stated in \cite{Bhattacharya_2017,HKLL_2006, Sorokhaibam:2019qho}. The entanglement entropy of the unmeasured fermions (when $N-M\ll N$) is dominated by the entropy contribution of the ground state. This is proportional to the size of this subsystem \cite{Zhang_2020,Sachdev_SEE_2016, Liu_SEE_2018}. 
One finds that the entropy of such a subsystem is given by
\begin{equation}\label{eq: S_UV}
    S_{\text{UV}} = -\log\left(\left(\frac{1}{2} + \mathcal{Q}\right)^2 + \left(\frac{1}{2}-\mathcal{Q}\right)^2\right).
\end{equation}
Here, the expression for $S_{\text{UV}}$ was found in \cite{Zhang_2020} for the case when one of the subsystems (in our case the unmeasured fermion subsystem) is sufficiently smaller than the other. The transition of the QES from the bifurcate surface to close to the asymptotic boundary is called the entanglement wedge transition \cite{Antonini_2023}.

\subsection{Comparison to entanglement entropy} \label{sec:jt/syk}
\begin{figure}
    \centering
    \includegraphics{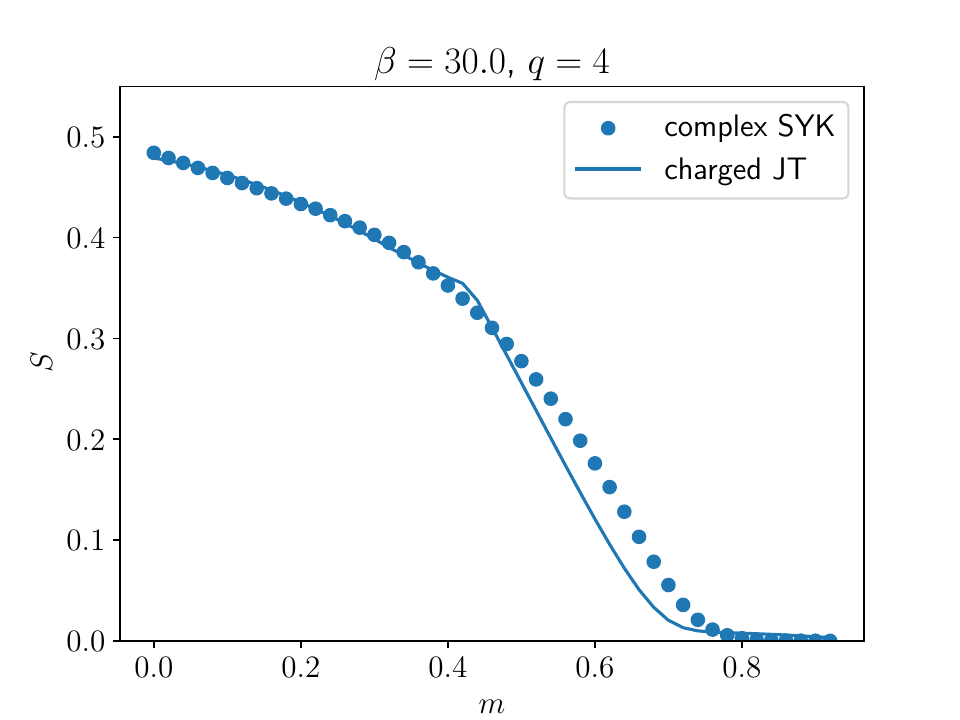}
    \caption{Comparison of the von Neumann entropy in the charged JT with part of the bulk cut off by an end of the world brane as derived in \ref{sec:bulk} and the single side R\'{e}nyi-2  entropy of the TFD state of the complex SYK after $M$ measurements as explained in \ref{section_syk_measurement}. $m=M/N$ is the ratio of measured to unmeasured fermions. Parameters are $\beta=30$ and $q=4$. The two results are in good agreement. Like \cite{Antonini_2023}, we find that the phase transition in the bulk is sharp, while it is more smoothed out in the boundary. However, the shape of the curve is qualitatively different from the result in \cite{Antonini_2023}. Notice that the curve does not extend to $m=1$. This is due to numerical instability around this point. However, one can easily extrapolate both the curves.}
    \label{fig:result}
\end{figure}

We can now turn to the comparison of the bulk entropy to the entanglement entropy of the TFD calculated in \cref{section_syk_measurement}. In \cref{fig:result}, we can see the numerical cSYK entanglement entropy $S_R$ as a function of measured fermions $m$ laid over the bulk entropy calculation (we gave $S_\text{Gen}$ an overall shift, to adjust for the offset at small $\mathcal{Q}$ we get from using Rényi-2 entropy, compare \cref{sec:neumannvrenyi}, $S_\text{UV}$ was left unchanged). The entanglement entropy for the cSYK and the bulk entropy for the charged JT decrease as the number of fermions being measured increases. Within this range both the models undergo a phase transition. After the phase transition, the number of permitted states is few and thus the entropy is low. The unmeasured fermions are already with a high probability in the positive charge eigenstate (see \cref{sec:charge_after_measurement,app: Phase structure}).

By \cite{cao2021thermodynamic,witten2020deformationsjtgravityphase}, the gaseous and liquid phases in the CFT are dual to large and small black holes respectively in the bulk. The intermediate unstable phase is dual to a medium size black hole. This tells us that our expression for $S_\text{Gen}$ is not applicable anymore after a certain charge is reached. Now, in principle there are two transitions: the entanglement wedge transition and the thermodynamic phase transition. However, we perform a minimisation procedure between $S_{\text{Gen}}$ and $S_{\text{UV}}$ only, such that
\begin{equation}
    S_{\text{QES}} = \min\{S_{\text{Gen}},S_{\text{UV}}\}.
\end{equation}
  We find that the entropy curve is already well described by the $S_\text{UV}$ formula for large charges. From $M_*\approx 0.4N$ onwards the minimal entropy is given by $S_{\text{UV}}$. We do not expect significant changes by including the thermodynamic phase transition.

The bulk teleportation procedure can be understood in a similar manner as in \cite{Antonini_2023,Antonini_2022}. Again, we consider that the measurement is performed on the left boundary.  Prior to any measurement being performed, the entanglement wedge of the left side is one entire side of the two sided black hole. Thus, the QES is located at the horizon and the entire bulk information is contained in the unmeasured fermions. While $M<M_*$, the entanglement wedge is cut off by the presence of the ETW brane for the bulk fermions dual to the $M$ measured fermions on the left side. However, the entanglement wedge of the left side still contains the centre of the bulk. Therefore, the bulk information is still contained in the $N-M$ unmeasured boundary fermions. In a sense, the bulk information gets teleported from the measured to the unmeasured fermions in the same side \cite{Antonini_2023}. Once $M>M_*$, the entanglement wedge for the left side sits at the UV cut off near the left asymptotic boundary. Simultaneously, the entanglement wedge of right side now extends all the way to the left asymptotic boundary and also contains a part of the interior of the black hole. In this fashion, the bulk information (except a small wedge near the cut off of the boundary) is now teleported from the left side to the other boundary \cite{Antonini_2023}. In the following section, we support this notion by formulating a teleportation protocol.

\section{Quantum Teleportation}
\label{sec:teleportation}

The authors of \cite{Antonini_2023} devised a teleportation protocol that serves in sending quantum information from one side of the TFD to the other (similar procedures were developed in \cite{Nezami_2023,gao2021traversable,Qi_2019,Mori:2024gwe}). In this section, we will adapt the teleportation protocol to the Dirac case to see if the charged wormhole becomes traversable as well.

\subsection{Teleportation protocol}\label{sec: Teleportaition protocol}
Before starting with our calculations, we first  give some definitions and derive a few useful relations. This will be mostly analogous to the Majorana case and we will thus follow \cite{Antonini_2023}. Likewise, we shall consider the case where all fermions are measured\footnote{Consequently, the entanglement wedge of the right side of the TFD will contain almost all of the entire bulk \cite{Antonini_2023}, meaning that the bulk can be reconstructed from sole knowledge of the right side.}.

The teleportation sequence is as follows. At $t=-t_0$ in Lorentzian time, the operator to be teleported $S$, e.g.\ a string of fermions, is inserted in the left side. At $t=0$, the measurement is performed in the left side, while simultaneously in the right side a decoding operator $D$ is applied. Finally, at $t=t_0$ the string $S$ is teleported to the right side \cite{Antonini_2023}. Here, the decoding operator $D$ is defined as follows
\begin{equation}\label{eq: decoding op}
    D \equiv \exp\left(-i\frac{\theta}{q}\sum_{k=1}^N c_kc_k^{\dagger}\right),
\end{equation}
where $\theta$ is a tuning parameter. Moreover, the combined action of measuring and decoding at the instant $t=0$ creates a quantum channel associated with the Kraus operator $K$ given by
\begin{equation}
    K\equiv(\ketbra{L})\otimes D.
\end{equation}
The whole procedure is subsumed in the left-right correlation function \cite{Antonini_2023}
\begin{equation}
    C_S = \frac{\bra{TFD} S_R(t_0) K S_L(-t_0) \ket{TFD}}{\bra{TFD} (\ketbra{L}{L})\otimes\mathbb{I} \ket{TFD}}. \label{eq:left-right_correlator}
\end{equation}
The Heisenberg picture version of $S$ is  defined via the time evolution operator, as $S_L(-t_0) = [U_L(t_0)SU_L^{\dagger}(t_0)]\otimes \mathbb{I}$ and $S_R(t_0) = \mathbb{I}\otimes[U_R^{\dagger}(t_0)SU_R(t_0)] $ respectively. We shall require $SS^{\dagger}=1$, so that $C_S \leq 1$. The denominator of \cref{eq:left-right_correlator} is solely there to remove the probability for the measurement outcome ($\ket{\uparrow\dots\uparrow}$). It reduces to $e^{-\beta E_{Q_\text{max.}}}$. \\
\\
As discussed in \cite{Antonini_2023}, the left-right correlator $C_S$ is closely related to the teleportation fidelity. Its calculation will be the goal for the rest of this section. To this end, we will take a slight detour and first consider the twisted correlation function, defined as
\begin{equation}
    G_S = \frac{\bra{TFD} S_R(t_0) K S_L(-t_0) \ket{TFD}}{\bra{TFD} K \ket{TFD}}. \label{eq:twisted_correlator}
\end{equation}
As it turns out, correlation functions for more complex strings $S$ can approximately be reconstructed from the basic correlation function, i.e.\ the correlation function for the case where only one qubit is sent \cite{Antonini_2023}. Therefore for simplicity, let $S_{L}\equiv (c_{L,k} + c_{L,k}^{\dagger})$ and $S_{R}\equiv \eta e^{-i \Delta \phi(Q)}(c_{R,k} + c_{R,k}^{\dagger})$, where the phase is required by \cref{eq:Theta}.

Before calculating the twisted correlation function (\ref{eq:twisted_correlator}), we first derive a useful property of the thermofield double. Taking into account the presence of $\Theta$ (see \cref{TFD_cSYK}), we can take operators from the left side of the TFD to the right via the relation
\begin{equation}
    O_\text{L}\ket{\infty}=\Theta O_\text{R}^{\
}\Theta^{-1}\ket{\infty}.\label{eq:tfd_left_to_right}
\end{equation}
To see that this is indeed true, we construct, for any pair of operators $O_\text{L}$ and $O_\text{R}$, an operator $\Tilde{O} = \left(O_\text{L} - \Theta O_\text{R}^{\dagger}\Theta^{-1}\right)e^{\beta(H_\text{L}+H_\text{R})}$ \cite{Cottrell_2019} and act with it on the TFD state. By inserting the identity ${\mathbb I_Q} = \sum_{m,m^{'}}\ketbra{m_Q}_L\otimes (\ketbra{\Theta m^{'}_Q}_R)$, we get
\begin{align}
    \Tilde{O}\ket{\text{TFD}} &\propto \sum_{Q=-N}^{N} e^{ -\mu Q} \sum_{n_Q, m_Q}  \left[ \braket{m_Q, On_Q}\ket{m_Q}_L \otimes \Theta \ket{n_Q}_R\right. 
    \nonumber
    \\
    &- \left.\braket{\Theta m_Q , \Theta O^\dagger n_Q}\ket{n_Q}_L \otimes \Theta \ket{m_Q}_R\right].
\end{align}
We can now use the anti-unitarity of $\Theta$ to obtain 
\begin{equation}
\braket{\Theta m_Q , \Theta O^\dagger n_Q}= \braket{m_Q,  O^{\dagger} n_Q}^* = \braket{ n_Q, O m_Q}
\end{equation}
and after renaming $n$ to $m$ in the second term the proposition \cref{eq:tfd_left_to_right} follows. We want the TFD state to be neutral in charge initially (when no measurement has been done yet). It is easy to see that this is only possible if $\mu$ vanishes. Thus from here on, we will set $\mu =0$. Whenever this is the case, the Hamiltonian is particle-hole symmetric and we can simply choose $\Theta$ proportional to charge conjugation $\mathcal{C}$, with $\mathcal{C}^{-1} Q \mathcal{C} = -Q$ and $\mathcal{C}^{-1}=\mathcal{C}^\dagger=(-1)^{N(N-1)/2}\mathcal{C}$\footnote{For an anti-unitary operator $A$ the definition of the adjoint changes to $\braket{Ax, y}=\braket{x, A^{\dagger}y}^*$.}, which implies $\mathcal{C}^{-1} c_i \mathcal{C} = \eta c_i^{\dagger}$, where $\eta(N) = \pm 1$. The  factor of proportionality is given by a phase that depends on $Q$, i.e.\ \cite{Sahoo_2020}
\begin{align}
    \Theta = e^{i\phi(Q)}\mathcal{C}. \label{eq:Theta}
\end{align}
We can use this property, to rewrite the numerator of \cref{eq:twisted_correlator} as follows
\begin{align}
    \text{Num.} &= \bra{\infty} e^{-\beta H_\text{L}/2} S_R(t_0) K S_L(-t_0) e^{-\beta H_\text{L}/2} \ket{\infty} \nonumber\\
    &= \bra{\infty} S_L(-t_0) e^{-\beta H_\text{L}/2} K S_L(-t_0) e^{-\beta H_\text{L}/2} \ket{\infty} \nonumber \\
    &= \sum_{n,Q} \bra{E_n, Q} e^{i\frac{\theta}{q}\sum c_{k, \text{L}}c_{k, \text{L}}^{\dagger}} S_L(-t_0) e^{-\beta H_\text{L}/2} \ketbra{L} S_L(-t_0) e^{-\beta H_\text{L}/2} \ket{E_n, Q} \nonumber \\
    &=  \sum_{n,Q}\bra{L} S_L(-t_0) e^{-\beta H_\text{L}/2} \ketbra{E_n, Q}_\text{L} e^{-i\frac{\theta}{q}\sum c_{k, \text{L}}c_{k, \text{L}}^{\dagger}} S_L(-t_0) e^{-\beta H_\text{L}/2}\ket{L}\nonumber \\
    &= \bra{L} S_L(-t_0) e^{-\beta H_\text{L}/2} e^{-i\frac{\theta}{q}\sum c_{k, \text{L}}c_{k, \text{L}}^{\dagger}} S_L(-t_0) e^{-\beta H_\text{L}/2}\ket{L}.
\end{align}
Similarly, for the denominator one finds
\begin{align}
    \text{Den.} = \bra{L} e^{-\beta H_\text{L}/2} e^{-i\frac{\theta}{q}\sum c_{k, \text{L}}c_{k, \text{L}}^{\dagger}} e^{-\beta H_\text{L}/2}\ket{L}.
\end{align}
Since now all operators have been brought to the left side and the right side has been traced out, we will drop the subscript L from here on. In conclusion, we now have
\begin{equation}
    G_S = \frac{\bra{L}e^{-\beta H} S(-t_0+i\beta)  e^{-i\frac{\theta}{q}\sum c_{k}(i\beta/2)c_{k}^{\dagger}(i\beta/2)} S(-t_0+i\beta/2) \ket{L}}{ \bra{L} e^{-\beta H} e^{-i\frac{\theta}{q}\sum c_{k}(i\beta/2)c_{k}^{\dagger}(i\beta/2)} \ket{L} }.
\end{equation}
After going to imaginary time, we can subsequently define the twisted correlator for general time arguments $\tau_1$, $\tau_2$ as
\begin{equation}
    G_S(\tau_1, \tau_2) = \frac{\bra{L}e^{-\beta H} T\left( e^{-i\frac{\theta}{q}\sum c_{k}(i\beta/2)c_{k}^{\dagger}(i\beta/2)} S(\tau_1) S(\tau_2) \right)\ket{L}}{ \bra{L} e^{-\beta H}T\left( e^{-i\frac{\theta}{q}\sum c_{k}(i\beta/2)c_{k}^{\dagger}(i\beta/2)} \right)\ket{L} }.
\end{equation}
We recover our original definition of $G_S$ by setting $\tau_1 = i\beta - t_0$, $\tau_2 = i\beta/2 - t_0$. $C_S$ in (\ref{eq:left-right_correlator}) is then proportional to $G_S$ via a factor $\Psi$, defined as \cite{Antonini_2023}
\begin{align}
    \Psi &= \frac{\bra{TFD} K \ket{TFD}}{\bra{TFD} (\ketbra{L}{L})\otimes\mathbb{I} \ket{TFD}}\\
    &= \exp\left[-\theta\frac{i}{q}\frac{\bra{L}e^{-\beta H} T\left( e^{-i\frac{\theta}{q}\sum c_{k}(i\beta/2)c_{k}^{\dagger}(i\beta/2)} \sum_j c_j(i\beta/2) c_j^{\dagger}(i\beta/2) \right)\ket{L}}{ \bra{L} e^{-\beta H} T\left( e^{-i\frac{\theta}{q}\sum c_{k}(i\beta/2)c_{k}^{\dagger}(i\beta/2)} \right)\ket{L} }\right].
\end{align}
Similar to the case for the measurement operator insertion discussed in \cref{section_syk_measurement}, the effect of the decoding operator's presence can be re-expressed in terms of boundary conditions imposed on the path integral. For later convenience, we define the field
\begin{equation}\label{eq: non-replicated chi}
    \chi_k(s) = \frac{i}{\sqrt{2}}\begin{cases}
        (c_k^{\dagger}-c_k)(is) & 0 < s < ,\beta\\
        (c_k^{\dagger}+c_k)(i2\beta-is) & \beta < s < 2\beta .
    \end{cases}
\end{equation}
The propagator 
\begin{equation}
    G_\chi(\tau_1, \tau_2) = \frac{1}{N} \sum_i \chi_i(\tau_1)\chi_i(\tau_2),
\end{equation}
denotes the correlator in the presence of the twisted boundary conditions. Using this we can express $\Psi$ as
\begin{eqnarray}
    \Psi &=&\exp\left\{-\theta\frac{iN}{4q}\left[  G_\chi(\beta/2, \beta/2) - G_\chi(3\beta/2, 3\beta/2) \right.\right. \nonumber\\ & &
    \left. \left. + G_\chi(\beta/2, 3\beta/2) - G_\chi(3\beta/2, \beta/2)\right]
    \vphantom{\frac{A}{B}}\right\},
\label{eq: ratio of G_chi and C_S}
\end{eqnarray}
where the aforementioned boundary conditions need to be imposed on $G_\chi$.

\begin{figure}
    \begin{subfigure}{0.5\textwidth}
        \centering
        \includegraphics[width=\textwidth]{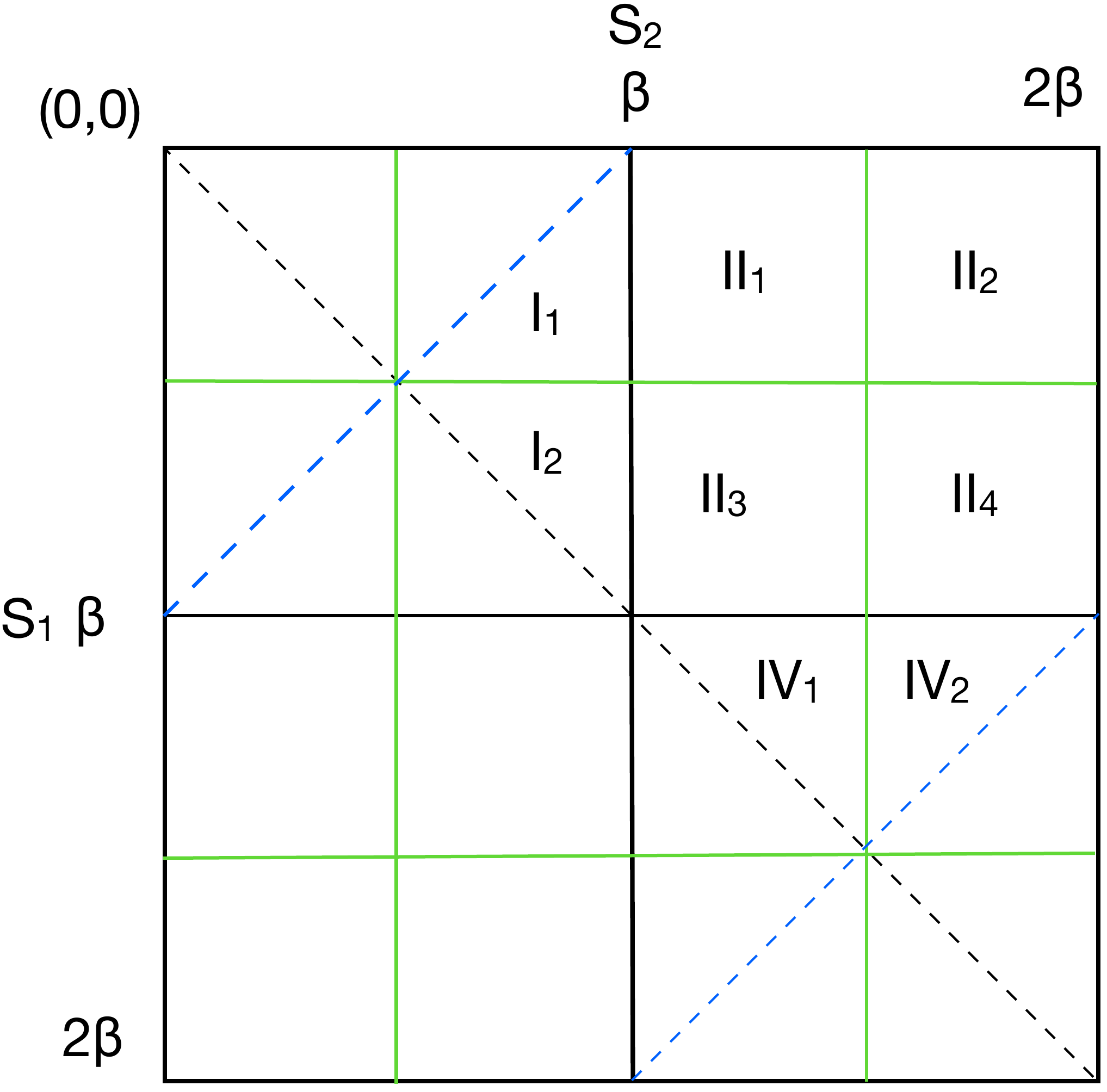}
        \caption{}
        \label{fig: fundamental regions}
    \end{subfigure}
    \begin{subfigure}{0.5\textwidth}
        \centering
        \includegraphics[width = \textwidth]{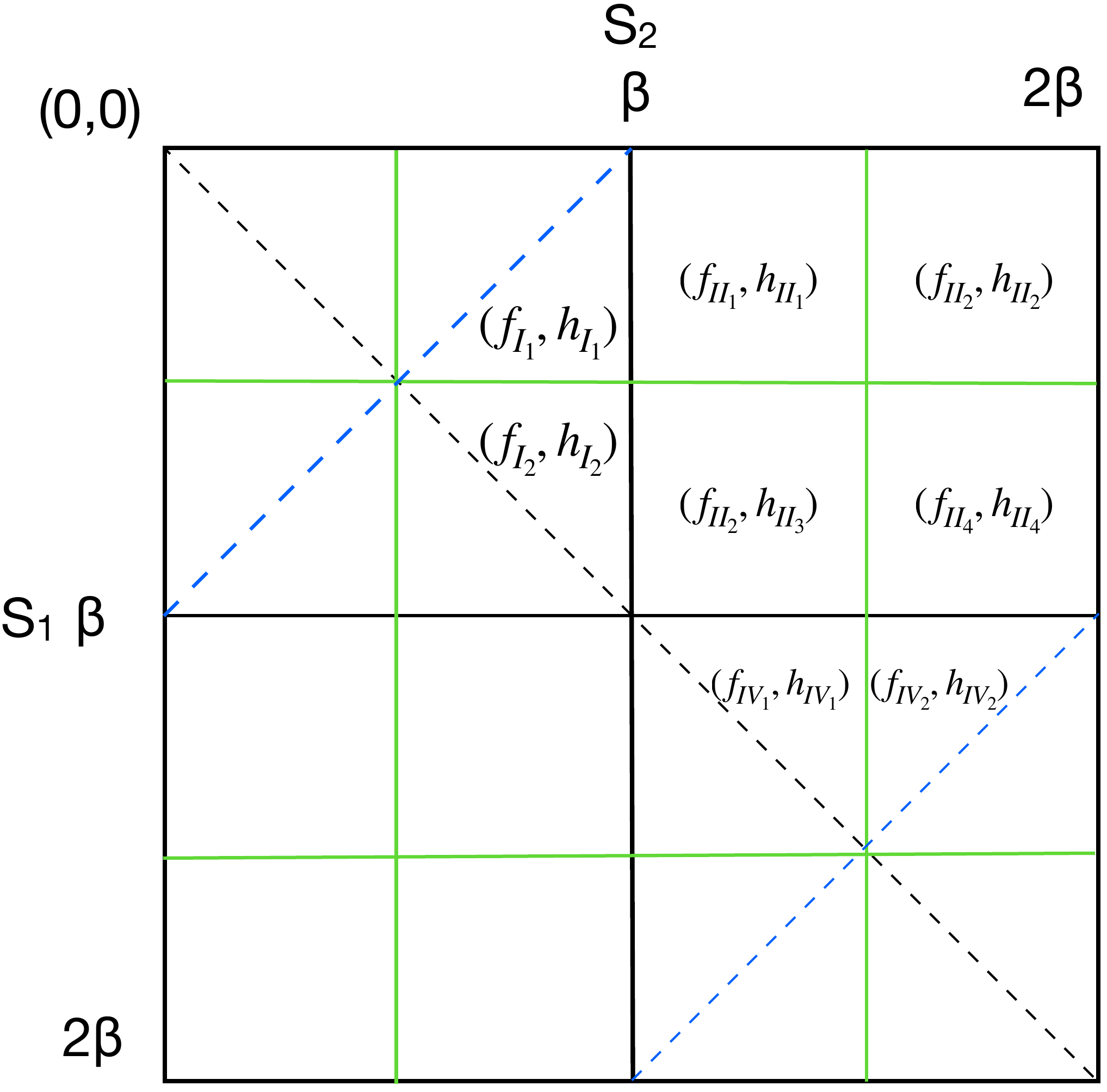}
        \caption{}
        \label{fig: functions in fundamental regions}
    \end{subfigure}
    \caption{The dashed red and blue lines indicate the anti-reflection and the reflection boundary conditions respectively. The solid orange/yellow lines indicate the twisted boundary conditions due to the decoding operator. $\text{I}_2$ ,$\text{II}_3$, $\text{IV}_1$ and $\text{II}_2$ are subregions with translation invariance, while $\text{I}_1$, $\text{II}_1$, $\text{II}_4$ and $\text{IV}_2$ do not have translation invariance. (b) $f_i,h_i$ are the functions used to express the general solution of the two dimensional Liouville equation.}
\end{figure}

Next, we focus on finding the analytic solution to the Schwinger-Dyson equation for $G_\chi$ in the large $q$ limit. Using this we can obtain a closed expression for $\Psi$ via \eqref{eq: ratio of G_chi and C_S}. In terms of the $\chi$ field, we can express all the correlation functions and the self-energies as
\begin{align}
    \hat{G} &= -\frac{1}{2}\begin{bmatrix}
        G_{11}(s_1,s_2) & G_{12}(s_1,2\beta-s_2)\\
        G_{21}(2\beta-s_1,s_2) & G_{22}(2\beta-s_1,2\beta-s_2)
    \end{bmatrix},\\
    \hat{\Sigma} &= -\begin{bmatrix}
        \Sigma_{11}(s_1,s_2) & \Sigma_{21}(s_1,2\beta-s_2)\\
        \Sigma_{12}(2\beta-s_1,s2) & \Sigma_{22}(2\beta-s_1,2\beta-s_2)
    \end{bmatrix}.
\end{align}
Using this, we can recast the Schwinger-Dyson equations as
\begin{align}
    &\hat{G} = (\partial-\hat{\Sigma})^{-1}, & \hat{\Sigma}(s_1,s_2) = J^2\hat{G}(s_1,s_2)^{q/2}\hat{G}^*(s_1,s_2)^{q/2-1}.
\end{align}
Next, we re-express the boundary conditions in terms of $\hat{G}$ as
\begin{align}
    \hat{G}(s,s) &= \frac{1}{2}, \:\:\:\:\:\hat{G}(s_1+2\beta,s_2) = -\hat{G}(s_1,s_2)\nonumber \\
    \hat{G}(s_1,s_2)^{\dagger} &=-\frac{1}{2} \begin{bmatrix}
        G_{11}(-s_2,-s_1) & G_{12}(-s_2,-2\beta+s_1)\\
        G_{21}(-2\beta+s_2,-s_1) & G_{22}(-2\beta+s_2,-2\beta+s_1)
    \end{bmatrix}.
\end{align}
Where we have used 
\begin{align*}
  [G_{ij}(s_1,s_2)]^{\dagger} = \begin{cases}
      G_{ii}(-s_2,-s_1) &\text{for} \;\;\;i=1,2\\
      -G_{ji}(-s_2,-s_1) &\text{for}\;\;\; i\neq j
  \end{cases}  
\end{align*}
to obtain the final relation.\\
To solve the Schwinger-Dyson equations analytically, we make use of the large $q$ limit, where we make the ansatz, $\hat{G}(s_1,s_2) = \hat{G}_0(s_1,s_2)\left(1+\frac{g(s_1,s_2)}{q}+\cdots\right)$ such that higher order terms are sufficiently suppressed. We obtain
\begin{equation}\label{eq: large q G_0}
    \hat{G}_0(s_1,s_2) = -\frac{1}{2}\begin{bmatrix}
        \text{sgn}(s_1-s_2) & -1\\
        1 & \text{sgn}(s_1-s_2)
    \end{bmatrix}.
\end{equation}
This leads to a Liouville equation for $g(s_1,s_2)$ \cite{Davison_2017, Bhattacharya_2017, gao2021traversable, Qi_2019, Eberlein_2017}
\begin{equation}\label{eq: Lioville eq for g}
    \partial_{s_1}\partial_{s_2}\left[\hat{G}_0(s_1,s_2)g(s_1,s_2)\right] = 2\mathcal{J}^2\hat{G}_0(s_1,s_2)\text{e}^{\frac{1}{2}\left[g(s_1,s_2)+g(s_2,s_1)\right]}.
\end{equation}
Rather than solving \eqref{eq: Lioville eq for g} in the different subregions produced due to the presence of the decoding operator, we will first find the fundamental region by use of various symmetries and then solve the Liouville equation in this region. In order to calculate the twisted correlation function we are interested in the case where all the fermions have been measured i.e.\ $m=1$. From the discussion given in \cref{app: Check for translation invar}, we see that the system has translation invariance when $m=1$. However, the introduction of the decoding operator \eqref{eq: decoding op} at $\tau = \beta/2$ breaks translation invariance in certain regions. This is because the decoding operator introduces twisted boundary conditions for which as $\tau \rightarrow \frac{\beta^+}{2}$, the fermions are expressed as a linear combination of both $c_k+c^{\dagger}_k$ and $c_k-c^{\dagger}_k$. The twisted boundary conditions are \footnote{These are obtained by conjugating $c_k+c^{\dagger}_k$ and $c_k-c^{\dagger}_k$ with the decoding operator.}
\begin{equation}
   \lim_{\tau \rightarrow \beta^+/2}\begin{pmatrix}
       c_k+c^{\dagger}_k\\
       c_k-c^{\dagger}_k
   \end{pmatrix} = \begin{pmatrix}
       \cos{\frac{\theta}{q}} & -i\sin{\frac{\theta}{q}}\\
       -i\sin{\frac{\theta}{q}} & \cos{\frac{\theta}{q}}
   \end{pmatrix}
   \lim_{\tau \rightarrow \beta^-/2}
   \begin{pmatrix}
       c_k+c^{\dagger}_k\\
       c_k-c^{\dagger}_k
   \end{pmatrix}.
\end{equation}
Using this, we can express the boundary conditions for $G_{\chi}(s_1,s_2)$ at $\tau = \beta/2$ as
\begin{equation}
    \begin{pmatrix}
       \lim_{s_{1/2} \rightarrow \beta^+/2}G_{\chi}(s_1,s_2)\\
       \lim_{s_{1/2} \rightarrow 3\beta^-/2}G_{\chi}(s_1,s_2)
   \end{pmatrix} = \begin{pmatrix}
       \cos{\frac{\theta}{q}} & -i\sin{\frac{\theta}{q}}\\
       -i\sin{\frac{\theta}{q}} & \cos{\frac{\theta}{q}}
   \end{pmatrix}
   \begin{pmatrix}
       \lim_{s_{1/2} \rightarrow \beta^-/2}G_{\chi}(s_1,s_2)\\
       \lim_{s_{1/2} \rightarrow 3\beta^+/2}G_{\chi}(s_1,s_2)
   \end{pmatrix}.
\end{equation}
In the large $q-$limit, this reduces to
\begin{align}\label{eq: large q twist b.c.}
    \begin{pmatrix}
       \lim_{s_{1/2} \rightarrow \beta^+/2}G_{\chi}(s_1,s_2)\\
       \lim_{s_{1/2} \rightarrow 3\beta^-/2}G_{\chi}(s_1,s_2)
   \end{pmatrix} &= \begin{pmatrix}
       1 & -i\frac{\theta}{q}\\
       -i\frac{\theta}{q} & 1
   \end{pmatrix}
   \begin{pmatrix}
       \lim_{s_{1/2} \rightarrow \beta^-/2}G_{\chi}(s_1,s_2)\\
       \lim_{s_{1/2} \rightarrow 3\beta^+/2}G_{\chi}(s_1,s_2)
   \end{pmatrix}\nonumber \\
   &\simeq \text{e}^{-i\frac{\theta}{q}}\begin{pmatrix}
       \lim_{s_{1/2} \rightarrow \beta^-/2}G_{\chi}(s_1,s_2)\\
       \lim_{s_{1/2} \rightarrow 3\beta^+/2}G_{\chi}(s_1,s_2)
   \end{pmatrix}.
\end{align}
We begin by making use of the the fact that for an operator $\mathcal{O}$
\begin{equation*}
    \mathcal{O}(\tau)^{\dagger} = [\text{e}^{H \tau}\mathcal{O} \text{e}^{-H \tau}]^{\dagger} = \mathcal{O}(-\tau),
\end{equation*}
to obtain the relations for $(c_i+c_i^{\dagger})(\tau)$ and $(c_i-c_i^{\dagger})(\tau)$ given as
\begin{align}\label{eq: dagger of Cs}
    [(c_i+c_i^{\dagger})(\tau)]^{\dagger} &= (c_i+c_i^{\dagger})(-\tau),\\
    [(c_i-c_i^{\dagger})(\tau)]^{\dagger} &= -(c_i-c_i^{\dagger})(-\tau).
\end{align}
Additionally, we also see that in order for the decoding operator and the twisted correlation functions to have real representations $\theta$ must be purely imaginary. This is done by first taking $\theta$ to be imaginary and then analytically continuing it to real values. The consequence of this is that $D\left(\frac{\beta}{2}\right)^{\dagger} = D\left(-\frac{\beta}{2}\right)$.
These results can be used along with the anti-periodicity of $\chi(s)$ to obtain
\begin{equation}\label{eq: reflection symm}
    G_{\chi}(s_1,s_2)^{\dagger} = G_{\chi}(\beta-s_2,\beta-s_1) = G_{\chi}(s_1,s_2).
\end{equation}
Where, the last relation is due to the fact that $G_{\chi}(s_1,s_2)$ has a real representation\footnote{We can conclude this as long as the initial value of $G$ is real (which in our case is taken to be $\frac{1}{2}\text{sgn}(\tau_1-\tau_2)$).}. Hence, the propagator $G_\chi\left( s_1 , s_2\right)$ is symmetric under reflections at  $s_1+s_2 = \beta$. Further, since we are considering $\mu = 0 \:\:$ and $ m=1 $, we also have the relation
\begin{equation}\label{eq: anti-reflection symm}
    G_{\chi}(s_1,s_2) = -G_{\chi}(s_2,s_1),
\end{equation}
i.e.\ anti-symmetry under reflections at $s_1=s_2$.
Eqs.\ \eqref{eq: reflection symm} and \eqref{eq: anti-reflection symm} can be used to reduce the full domain of the twisted correlation function down to the fundamental regions as shown in Figure \ref{fig: fundamental regions}. Employing \eqref{eq: anti-reflection symm}, we see that the Liouville equation for the cSYK model \eqref{eq: Lioville eq for g} becomes the same as the one for the real SYK because $g(s_1,s_2) = g(s_2,s_1)$.
Hence, for $G_{\chi}(s_1,s_2) = (G_{\chi})_0(s_1,s_2)\text{e}^{g(s_1,s_2)/q}$ we can express the twisted boundary conditions as
\begin{align}
    g\left(\frac{\beta^+}{2},s_2\right) + i\theta &= g\left(\frac{\beta^-}{2},s_2\right),\\
    g\left(s_1,\frac{3\beta^+}{2}\right) + i\theta &= g\left(s_1,\frac{3\beta^-}{2}\right).
\end{align}
Explicitly \cite{gao2021traversable,Qi_2019,Eberlein_2017} \footnote{The general solution of the two dimensional Liouville equation can be expressed as
\begin{equation*}
    g\left(s_1,s_2\right) = \log\left[\frac{f'(s_1)h'(s_2)}{(1-\mathcal{J}^2f(s_1)h(s_2))^2}\right].
\end{equation*}
The solution $g\left(s_1,s_2\right)$ is invariant under $PSL\left( 2,{\mathbb R}\right)$ transformations: $f(s)\mapsto\frac{a+b f(s)}{c + d f(s)}$ and $h(s)\mapsto \frac{d-c\mathcal{J}^2h(s)}{\mathcal{J}^2[-b+a\mathcal{J}^2h(s)]}$ with $\left( \begin{array}{cc} a & b\\ c & d\end{array}\right) \in SL\left( 2,{\mathbb R}\right)$.
}
\begin{align}
    \frac{f_i'\left(\frac{\beta}{2}^+\right)h_i'(s_2)\text{e}^{i\theta}}{\left(1+\mathcal{J}^2f_i\left(\frac{\beta}{2}^+\right)h_i(s_2)\right)^2} &= \frac{f_j'\left(\frac{\beta}{2}^-\right)h_j'(s_2)}{\left(1+\mathcal{J}^2f_j\left(\frac{\beta}{2}^-\right)h_j(s_2)\right)^2},\label{eq: twist b.c. in g A}\\
    \frac{f_i'(s_1)h_i'\left(\frac{3\beta}{2}^+\right)\text{e}^{i\theta}}{\left(1+\mathcal{J}^2f_i(s_1)h_i\left(\frac{3\beta}{2}^+\right)\right)^2} &= \frac{f_j'(s_1)h_j'\left(\frac{3\beta}{2}^-\right)}{\left(1+\mathcal{J}^2f_j(s_1)h_j\left(\frac{3\beta}{2}^-\right)\right)^2}\label{eq: twist b.c. in g B}.
\end{align}
By first integrating over $s_2$ in \eqref{eq: twist b.c. in g A} and over $s_1$ in \eqref{eq: twist b.c. in g B} followed by the use of the global $SL(2)$ symmetry of the general solution of \eqref{eq: Lioville eq for g} we can re-express the twist boundary conditions in the fundamental regions as
\begin{align}
    f_{\text{I}_1}\left(\frac{\beta}{2}\right) &= f_{\text{I}_2}\left(\frac{\beta}{2}\right), &f'_{\text{I}_1}\left(\frac{\beta}{2}\right) &= \text{e}^{i\theta}f'_{\text{I}_2}\left(\frac{\beta}{2}\right),\\
    h_{\text{I}_1}(\beta) &= h_{\text{IV}_1}(\beta),  &h'_{\text{I}_1}(\beta) &= h'_{\text{IV}_1}(\beta),\\ 
    h_{\text{IV}_2}\left(\frac{3\beta}{2}\right) &= h_{\text{IV}_1}\left(\frac{3\beta}{2}\right), &h'_{\text{IV}_2}\left(\frac{3\beta}{2}\right) &= \text{e}^{i\theta}h'_{\text{IV}_1}\left(\frac{3\beta}{2}\right),\\
    f_{\text{I}_1}(\beta) &= f_{\text{IV}_1}(\beta),  &f'_{\text{I}_1}(\beta) &= f'_{\text{IV}_1}(\beta). 
\end{align}
The global $SL(2)$ symmetry also allows us to transform $f_j$ and $h_j$ such that $f_i = f_j$. \Cref{fig: functions in fundamental regions} indicates the various subregions and the functions $f_i,h_i$ defined in these domains to obtain the expressions for $g(s_1,s_2)$.
Finally, we have to take care that in subregions $\text{I}_2$ and $\text{IV}_1$ for $s_1 = s_2$ we have
\begin{align}
\frac{f'(s_1)h'(s_2)}{(1-\mathcal{J}^2f(s_1)h(s_2))^2} = 1, \end{align}
as well as the case when $\theta = 0$ where we recover the usual correlator in the absence of the decoding operator.

\begin{figure}
    \centering
    \includegraphics{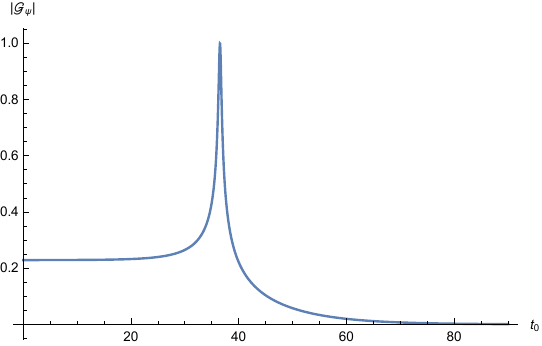}
    \caption{The magnitude of the twisted correlation function as a function of time. The plot is obtained for $\beta = 30,\:\: q = 4$. We have also used the low energy approximation for $\gamma = \frac{1}{\beta\mathcal{J}}+\mathcal{O}\left(\frac{1}{(\beta\mathcal{J})^2}\right)$}
    \label{fig:Twisted correlator plot}
\end{figure}

To determine the various $f_i,h_i$ in the subregions, we first make use of the translation invariance in subregions $\text{I}_2$ and $\text{IV}_1$ which implies that $f_{\text{I}_2} = h_{\text{I}_1} = f_{\text{IV}_1} = h_{\text{IV}_1} = \sin(\alpha\abs{s}+\gamma)$. The solution in these regions is then given by
\begin{equation}\label{eq: Liouville eq sol with translation invar}
    g(s) = 2\log\frac{\sin{\gamma}}{\sin(\alpha\abs{s_{12}}+\gamma)},
\end{equation}
for $\alpha = \mathcal{J}\sin{\gamma}$ and $\gamma = \frac{\pi - \mathcal{J}\beta\sin{\gamma}}{2}$.
The correlation function in the subregions which are not translation invariant are given by\footnote{We can use a similar solution as in \cite{Qi_2019, gao2021traversable} because although the SYK model with twisted boundary conditions has more constraints than the cSYK, both of them share certain constraints. Thus, the solutions given in \cite{Qi_2019, gao2021traversable} will also be a solution for the subregions in question, though not the most general one.}
\begin{equation}\label{eq: Liouville eq sol no translation invar}
    g(s_1,s_2) = 
     \log\frac{g(s)}{\left[1-\frac{1-\text{e}^{i\theta}}{\sin{\gamma}}\frac{\sin{(\alpha(s_1-\beta/2))}\sin{(\alpha(s_2-3\beta/2))}}{\sin{(\alpha(s_1-s_2)+\gamma)}}\right]^2} + i\theta.
\end{equation}
Using these solutions, we can calculate the expression for $\Psi$ in \eqref{eq: ratio of G_chi and C_S}. From the UV relations, we get that $G_{\chi}(\beta/2,\beta/2) = G_{\chi}(3\beta/2,3\beta/2) = 1/2$ while the anti-reflection symmetry gives $G_{\chi}(3\beta/2,\beta/2) = -G_{\chi}(\beta/2,3\beta/2) = -\frac{1}{2}\left[\frac{\sin{\gamma}}{\sin{(-\beta\alpha+\gamma)}}\right]^{2/q}$. Thus, $\Psi$ becomes
\begin{equation}\label{eq: expression for Psi}
    \Psi = \exp{\frac{i\theta N}{4q}\left(\frac{\sin{\gamma}}{\sin{(\gamma-\beta\alpha)}}\right)^{2/q}}.
\end{equation}
Hence, the twisted correlator and the left-right correlator are different up to a phase which is independent of the the choice of operator $S$ being teleported. Due to this, the two correlation functions have the same magnitude. This is the same as in the case of the real SYK model \cite{Antonini_2023}.
Finally, we find the expression for the twisted correlation function when $s_1 = \frac{\beta^-}{2}+it_0$ and $s_2 = \beta^- + it_0$
\begin{equation}
    G_{\text{I}_1}\left(\frac{\beta^-}{2}+it_0,\beta^-+it_0\right) \approx \text{e}^{i\theta/q}\left(\frac{\sin\gamma}{1-\frac{\theta\text{e}^{2\alpha t_0+i\gamma}}{4\sin\gamma}}\right)^{2/q}.
\end{equation}
Where we have assumed that $\theta\ll 1$ and $\text{e}^{-\alpha t_0} \ll 1$. The absolute value of $G_{\text{I}_1}$ is at a maximum when $\theta = 4\text{e}^{-2\alpha t_0}\sin\gamma$ and gives
\begin{equation}
    G_{\text{I}_1}\left(\frac{\beta^-}{2}+it_0,\beta^-+it_0\right) \approx \text{e}^{\frac{i\pi}{q}}\left(\cos\frac{\gamma}{2}\right)^{2/q}. 
\end{equation}
This is the same as the expression obtained for the twisted correlation function in \cite{Antonini_2023}. In Figure \ref{fig:Twisted correlator plot} we can see that the magnitude of the twisted correlator is initially constant, however after evolving for some time $t_0$ it quickly reaches the maximum and teleportation of a single operator can be successfully performed from one side to the other.

\newpage

\section{Conclusion}\label{sec:conclusion}

Inspired by \cite{Antonini_2023}, we studied the effect of adding a $U(1)$-charge measurement on one side of the thermofield double state in the complex SYK model. We post-selected samples in which the measurement returned only positive charges. Adding the measurements led to boundary conditions for the fermion fields, that needed to be taken care of when evaluating the path integral. By solving the Schwinger-Dyson equations numerically, we were able to compute the on-shell action and the entanglement entropy of either side of the TFD as a function of the fraction of measured fermions $m$. As expected, the entropy has its highest value at $m=0$ and then monotonically decreases. It already almost vanishes before all fermions are measured, in our example at around $m\approx0.8$. This can be explained by the fact, that the total charge of the fermionic system becomes maximal at this point. Hence, we concluded that the system exists in the gaseous phase when $m\ll0.7$ and in the liquid phase for $m\geq 0.8 $. When $0.7\leq m \leq 0.8$, the system undergoes a phase transition.

In \cref{section_bulk}, we studied the holographic dual of adding a measurement on the cSYK TFD. On the gravity side, we applied the quantum extremal surface formula to compute the entropy in a bulk region containing a gauge field coupled to gravity and $N-M$ or $M$ copies of a CFT or BCFT respectively. The JT entropy contribution was given by the dilaton value, the CFT/BCFT entropy was calculated to be constant (for fixed $m$) and the gauge field entropy contribution was computed by the saddle-point approximation. We approximated the location of the QES in the limits $M\ll N$ and $M\sim N$ and interpolated the entropy in between the two. The same qualitative behaviour as in the cSYK model could be observed by matching of the free parameters appearing in both theories. It was observed that, when $M<M_*$, the bulk information encoded within $M$ boundary fermions prior to the measurement got teleported to the $N-M$ unmeasured fermions on the same side after the measurement. On the other hand when $M>M_*$ fermions had been measured, the information was teleported to the other boundary. Furthermore, it was established that a single operator can be teleported from one side to the other via the formulation of an appropriate decoding operator.

For future work, it would be appealing to consider other post-selected measurement outcomes. For neutral charge, we expect the same outcome as for the real SYK. It would be interesting to find a critical charge per measurement value at which the phase transition value observed in this paper first occurs.  Moreover, we would like to add the small black hole phase to the minimisation procedure for the bulk entropy.

\section*{Acknowledgments}

We acknowledge support by
the Bonn Cologne Graduate School of Physics and Astronomy (BCGS) and the Deutsche Forschungsgemeinschaft (DFG) through Germany’s Excellence Strategy—Cluster of Excellence Matter and Light for Quantum Computing (ML4Q). Furthermore, Y.K.\ would like to thank the Rosa Luxemburg Foundation for the granted financial support. Finally, the authors gratefully acknowledge the granted access to the Bonna cluster hosted by the University of Bonn.
\newpage
\FloatBarrier
\appendix

\section{Phase Structure of the Complex SYK}\label{app: Phase structure}
In the following, we will give an overview of the phase structure of the complex SYK at $q=4$. This topic has already been discussed in detail in \cite{cao2021thermodynamic,Louw_2023,Garcia-Garcia:2020vyr}. Results reported in this appendix have been obtained by applying some of the techniques presented there.

\begin{figure}
    \centering
    \begin{subfigure}{0.449\textwidth}
        \includegraphics[width=\textwidth]{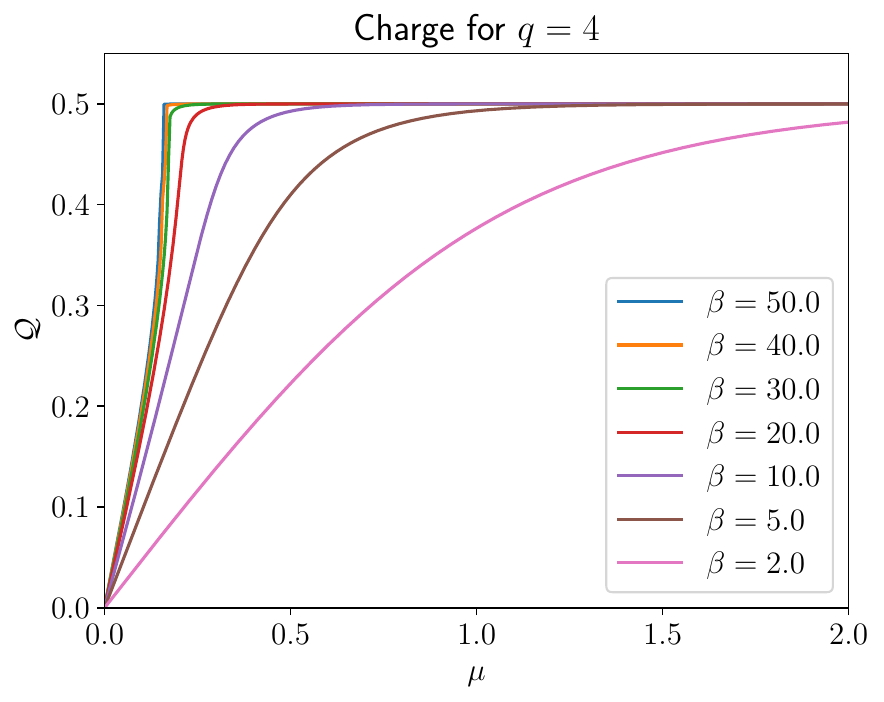}
        \subcaption{charge $\mathcal{Q}$}
    \end{subfigure}
    \hfill
    \begin{subfigure}{0.5\textwidth}
        \includegraphics[width=\textwidth]{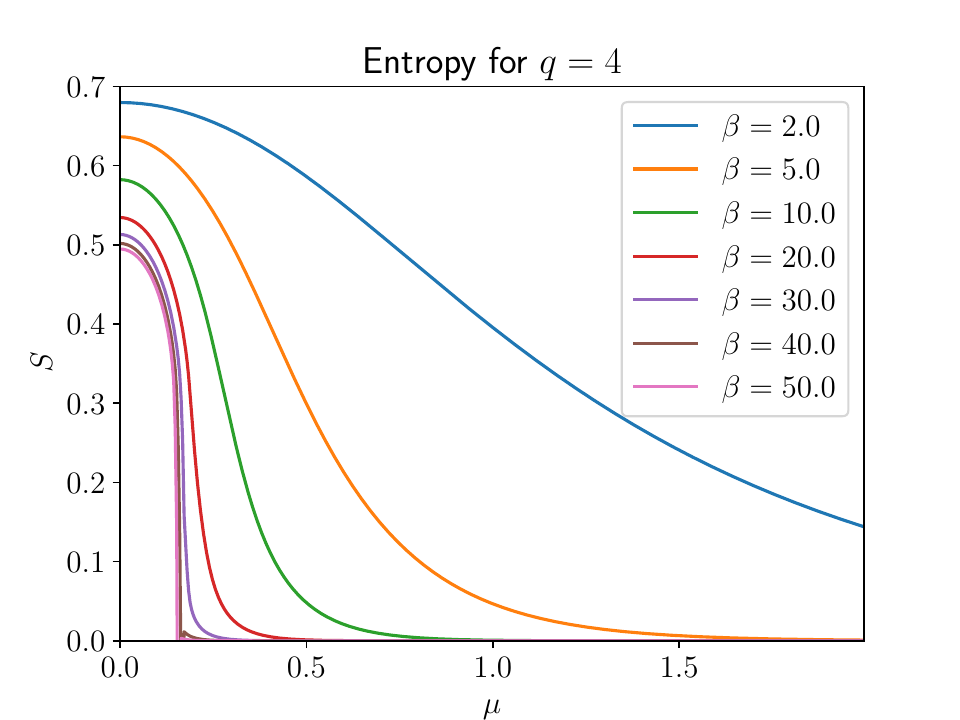}
        \subcaption{von Neumann Entropy $S$}
    \end{subfigure}
    \caption{Numerical results for charge $\mathcal{Q}$ and entropy $S$ as functions of the chemical potential $\mu$ for the cSYK in the grand canonical ensemble. We can see that for large $\mu$ the curves tend towards a common constant value. While this transition is smooth for high temperatures. There seems to be a sharp transition for low temperatures. This is due to a first order phase transition that takes place below some critical temperature $T_\text{crit.}$ \cite{cao2021thermodynamic}.}
    \label{fig:gc_results}
\end{figure}

We consider the cSYK in the grand canonical ensemble with a finite chemical potential $\mu$ and solve the Schwinger-Dyson equations iteratively\footnote{We use the Kitaev trick \mycite{Maldacena_2016}{introduced in} for updating the propagtor after each step, $G_\text{new} = (1-x)G_\text{old} - x(i\omega + \mu + \Sigma)^{-1}$, with $x\approx0.05$ and $x$ is halved every time the error grows.} in Fourier space
\begin{equation}
    G(\omega_j) \equiv \int_0^\beta d\tau e^{i\omega_j\tau} G(\tau), \; G(\tau) = \sum_{j=-\Lambda}^{\Lambda-1} e^{-i\omega_j\tau} G(\omega_j),
\end{equation}
here $\omega \equiv \pi\frac{2n+1}{\beta}$ and $\Lambda$ is some appropriate UV cut-off. We used the same notation and conventions as \cite{cao2021thermodynamic}. The iteration is stopped when the difference between the propagators of two consecutive iteration steps,
\begin{equation}
    \Delta G(\omega_j) = \frac{1}{\Lambda} \sum_{j=0}^{\Lambda-1} \left| G_\text{new}(\omega_j) - G_\text{old}(\omega_j)\right|,
\end{equation}
falls below a certain threshold, $\epsilon$. The on-shell action is then given by \cite{Davison_2017}
\begin{equation}
    I^{*} = -\frac{\mu\beta}{2} + \ln\left(1+e^{\mu\beta}\right)+ 2 \sum_{j=0}^{\Lambda-1} \left[ \ln{\left| \frac{i\omega_j+\mu + \Sigma(\omega_j)}{i\omega_j + \mu}\right|} + \left( 1 - \frac{1}{q}\right) \Re{G(\omega_j)\Sigma(\omega_j)} \right],
\end{equation}
from which we immediately get the grand potential
\begin{equation}
    \Omega = - \frac{I^{*}}{\beta}
\end{equation}
and the free energy $F$ is defined as its Legendre transform
\begin{equation}
    F = \Omega + \mu Q.
\end{equation}
Finally, the charge is calculated via
\begin{equation}
    \mathcal{Q} = -\frac{2}{\beta}\sum_{j=0}^{\Lambda-1}\Re{
    G(\omega_j)
    }.
\end{equation}
Our results have been obtained for a maximal error of $\epsilon = 10^{-14}$ and a cut-off of $\Lambda=2^{16}$.
\Cref{fig:gc_results} shows curves for the charge $\mathcal{Q}$ and the von Neumann entropy $S$\footnote{The procedure to compute the von Neumann entropy is described in \cref{sec:neumannvrenyi}.} as functions of $\mu$ for different values of the inverse temperature $\beta$. As we can see, both the $S$ and $Q$ curves asymptote to a constant. For large $\mu$, we get $\mathcal{Q} \rightarrow 1/2$ and $S\rightarrow 0$. While for small $\beta$ the transition is comparatively slow and smooth, it is noticeably sharper and more immediate for larger $\beta$. This is due to a first order phase transition that becomes possible below a certain temperature $T_\text{crit.}$ \cite{cao2021thermodynamic}. In the following, we shall refer to the two phases as liquid and gaseous phase. The two stable phases are separated, in $S\mathcal{Q}$-space, space, by an unstable region with negative specific heat, that is therefore never attained \cite{cao2021thermodynamic}.

Since in this paper, we mainly consider the low temperature limit, we are not interested in actually calculating $T_\text{crit.}$. Instead, we mainly wish to understand the behaviour of the system below criticality. To visualize the phase structure, we look at the grand potential $\Omega$ and the charge $\mathcal{Q}$ and compare them for the two phases. This is done by slowly varying the chemical potential from small to large (forward) and from large to small (backward), while for each step using the result from the previous step as a new initial guess for the propagator $G$. Below criticality, the system should then exhibit a hysteresis behaviour, therefore, in the region where the two phases coexist, allowing us to obtain two different results for the same $\mu$/$\mathcal{Q}$. 

\begin{figure}
    \centering
    \begin{subfigure}{0.49\textwidth}
        \includegraphics[width=\textwidth]{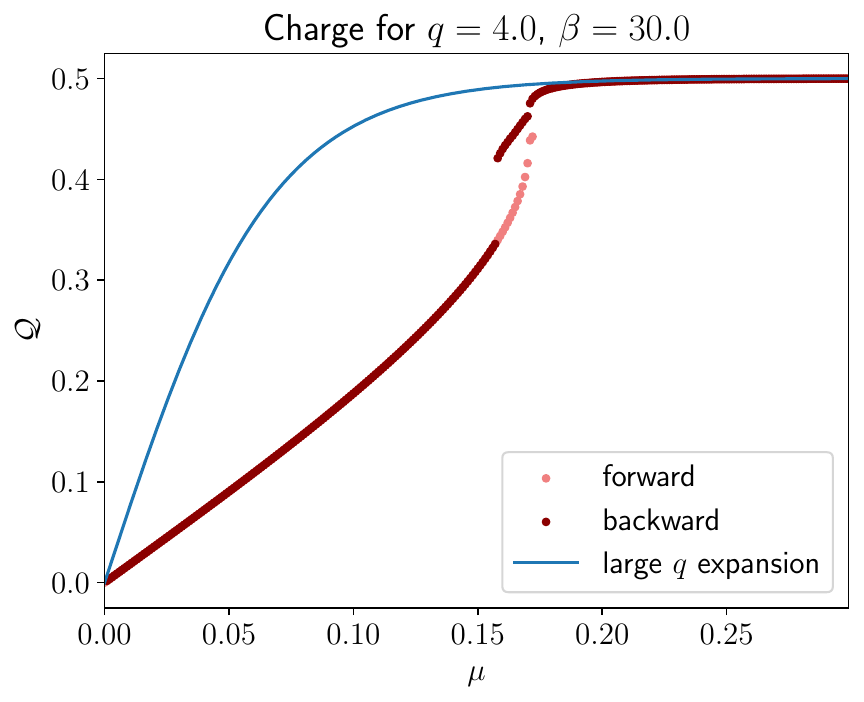}
        \subcaption{$\mathcal{Q}(\mu)$ at $\beta = 30$}
    \end{subfigure}
    \hfill
    \begin{subfigure}{0.49\textwidth}
        \includegraphics[width=\textwidth]{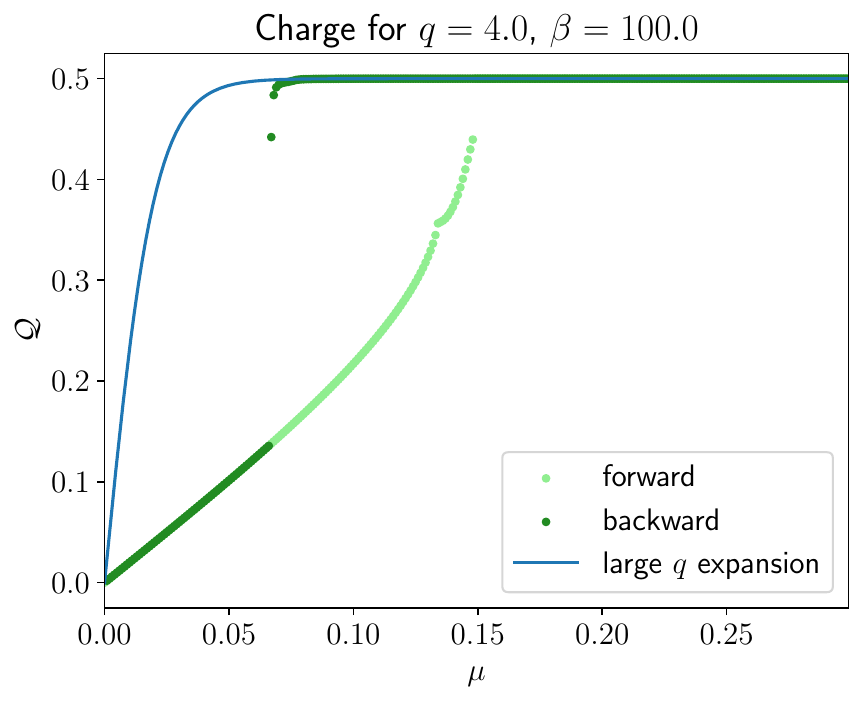}
        \subcaption{$\mathcal{Q}(\mu)$ at $\beta = 100$}
    \end{subfigure}
    \begin{subfigure}{0.49\textwidth}
        \includegraphics[width=\textwidth]{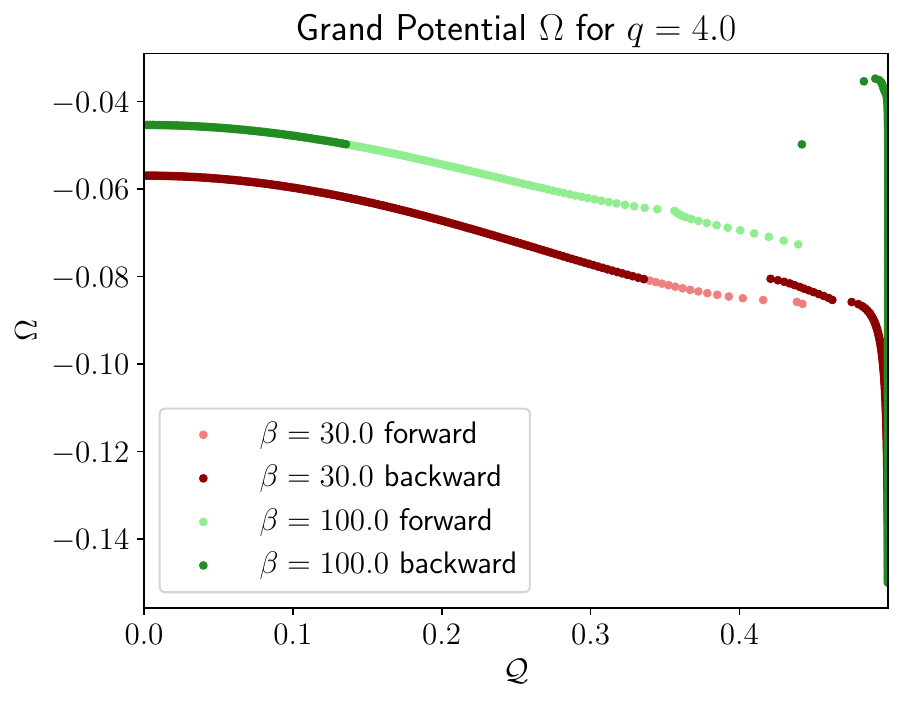}
        \subcaption{$\Omega(\mathcal{Q})$}
        \label{fig:  gc free energy as func of Q}
    \end{subfigure}
    \hfill
    \begin{subfigure}{0.49\textwidth}
        \includegraphics[width=\textwidth]{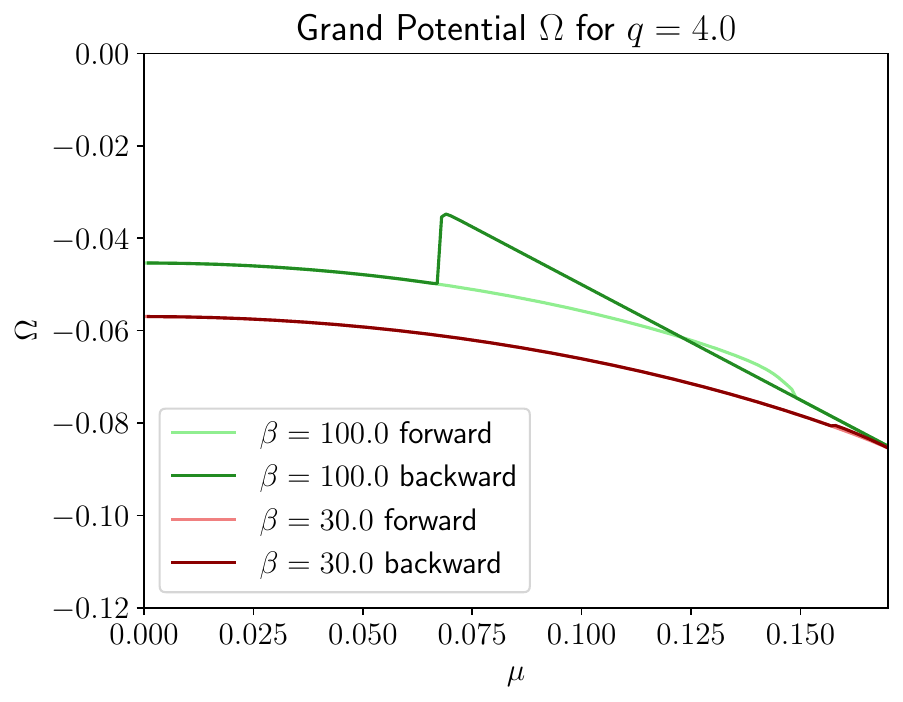}
        \subcaption{$\Omega(\mu)$}
    \end{subfigure}
    \caption{Charge $\mathcal{Q}$ and grand potential $\Omega$ from numerical computations with hysteresis. To get the forward curves, we slowly increased $\mu$ from $0$ to $0.3$, reusing the result for the propagator $G(\omega)$ of each point as initial guess for the next point. We get the backwards curves through equivalent means, but going from large to small $\mu$. This artificially keeps the system longer in the respective starting phase. The solid blue line is the large $q$ expansion to second order in $q^{-1}$}
    \label{fig:gc_free_energy}
\end{figure}

We give the corresponding plots for $\mathcal{Q}$ and $\Omega$ in \cref{fig:gc_free_energy} respectively, the plots are drawn at the same temperature as the one chosen for the JT/cSYK comparison in \cref{sec:bulk} ($\beta=30$) and an additional even lower temperature ($\beta=100$) where the effect is more pronounced. The curves show that there is only one stable solution to the Schwinger-Dyson equations for most values of $\mu$. However, there is a sliver of the $\mu$-axis where the forward and backward approach yield different charges. This is the region where both phases coexist and it increases with growing temperature, see also \cite{cao2021thermodynamic}. The hysteresis delays the phase transition and lets one of the phases enter the region of the other. The system eventually becomes overextended and rapidly reverts to the other phase. While for small $\mu$ the system is solidly in the gaseous phase and large $\mu$ puts in the liquid phase, phase transitions are generally possible anywhere in the region of coexistence. Nevertheless, if we let it evolve freely, the system will usually follow the line of least grand potential. 

Notice also that some values of the charge are never visited by neither the forward nor the backward curve. This is also reflected by the $\Omega(\mathcal{Q})$ curve\footnote{The $\Omega(\mathcal{Q})$ can also tell us more about where the system will end up after the phase transition has occurred. Basically, if the grand potential diverts too much from the thermodynamically favoured curve it will transition along a constant $\mu$ line through the $\Omega$-$\mathcal{Q}$ phase space until it hits the optimal curve again, skipping the intermediate $\mathcal{Q}$s.} being discontinuous. Even if we increase the number of points in $\mu$, the iteration does not seem to be able to converge to a stable solution within that region. This effect corresponds to the aforementioned thermodynamic instability, due to a would be negative specific heat \cite{cao2021thermodynamic}, which does not allow the corresponding configurations to get realized.

\begin{figure}
    \centering
    \includegraphics[width=\textwidth]{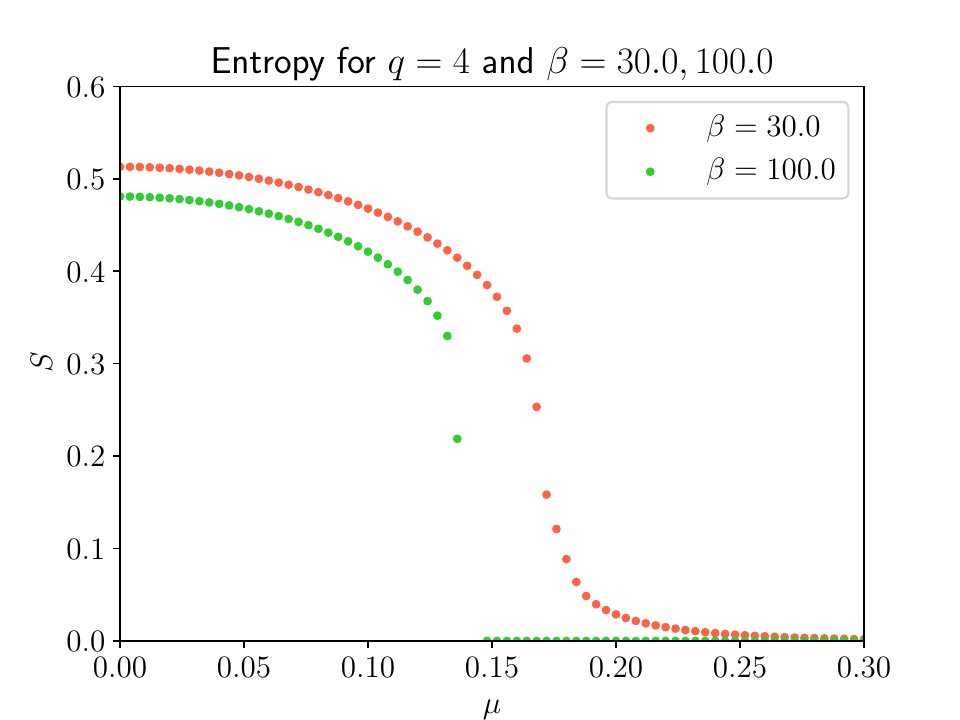}
    \caption{Numerical results for the von Neumann entropy $S$ as a function of the chemical potential $\mu$ for $\beta=30$ and $\beta=100$. We chose to plot points instead of lines hear, to show the discontinuous nature of the curves around the phase transition point. Some values of $S$ cannot be reached through a convergent iterative process for the respective value of $\beta$.}
    \label{fig:gc_entropy}
\end{figure}

Let us now conclude by considering entropy again. \Cref{fig:gc_entropy} shows the von Neumann entropy for $\beta=30$ and $\beta=100$. As was discussed above, initially the entropy varies slowly with $\mu$. Around the phase transition point in $\mu$, we then see it plummeting sharply and staying almost constant at close to zero in the liquid phase. We have learned from \cref{fig:gc_results,fig:gc_free_energy} that this is accompanied by the charge jumping to large values close to $0.5$. This explains why the entropy becomes so small in the liquid phase. At large absolute value of the total charge, the single fermion charges are almost all aligned and the charge subsectors at close to $\mathcal{Q}=0.5$ therefore contain very few states.
\begin{figure}
    \centering
    \includegraphics[width=\textwidth]{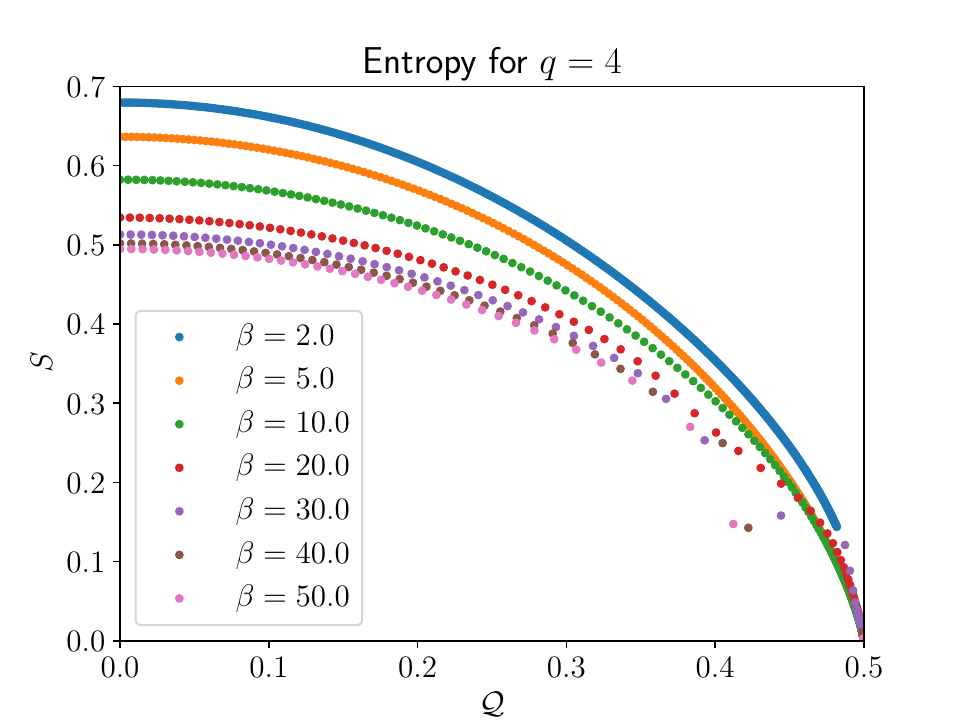}
    \caption{The von Neumann entropy in the grand canonical ensemble for various temperatures from numerical results. In our numerical calculations $\mathcal{Q}$ is a dependent variable and so points are not equidistant on the $x$-axis.}
    \label{fig:gc_results_v_charge}
\end{figure}

\section{Specific Heat and Charge Compressibility} \label{sec:K_and_gamma}

To do the JT/SYK comparison in \cref{sec:jt/syk}, we need certain thermodynamic quantities for the cSYK and match them to their JT counterparts (see \cref{theory_dual_cSYK}). For most of them either a closed expression exists (see in particular \cite{Gu_2020,Davison_2017}), or we can perform a large $q$ expansions. A large $q$ expansion for the specific heat $\gamma$ and the charge compressibility $K$ was derived in \cite{Davison_2017} up to second order in $q^{-1}$. However, as we can see in \cref{fig:gc_free_energy} at $q=4$ the system diverges significantly from the second order result. Unfortunately, deriving higher orders in the expansion is tricky and according to our knowledge closed forms do not exist. This is why we opt for a numerical approach (see \cref{app: Phase structure}). 
\begin{figure}[h]
    \centering
    \begin{subfigure}{0.49\textwidth}
        \includegraphics[width=\textwidth]{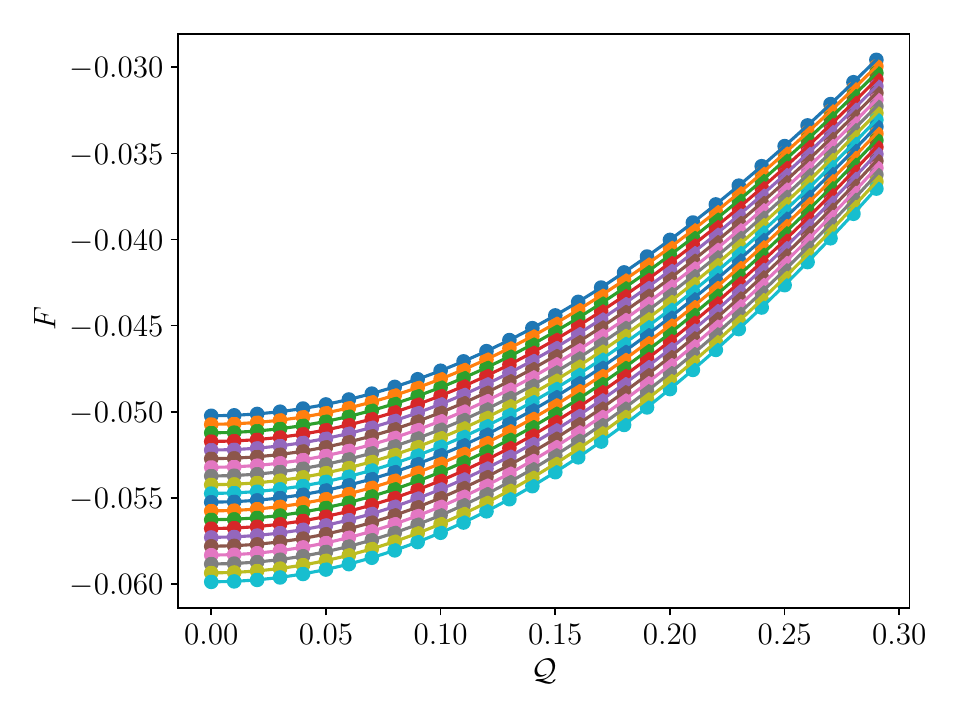}
    \end{subfigure} \hfill
    \begin{subfigure}{0.49\textwidth}
        \includegraphics[width=\textwidth]{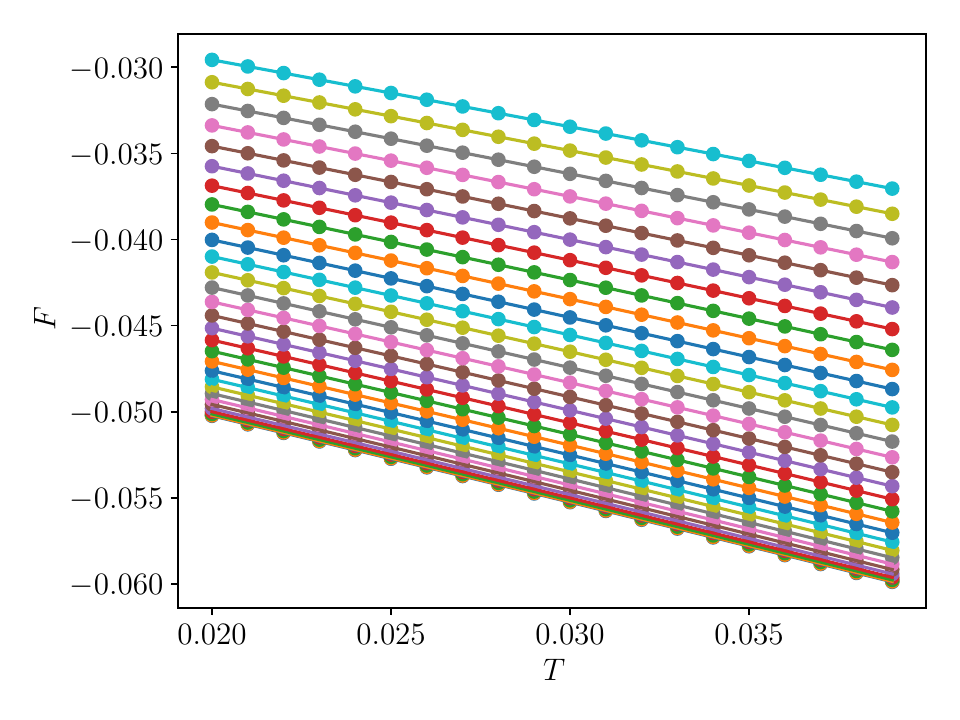}
    \end{subfigure}
    \caption{Constant $T$ and $\mathcal{Q}$ lines of the free energy $F$. Dots are numerical results from iterative solution of the Schwinger-Dyson equations. Solid lines are polynomial fits.}
    \label{fig:free_energy_beta_Q}
\end{figure}

\begin{figure}[h]
    \centering
    \begin{subfigure}{0.49\textwidth}
        \includegraphics[width=\textwidth]{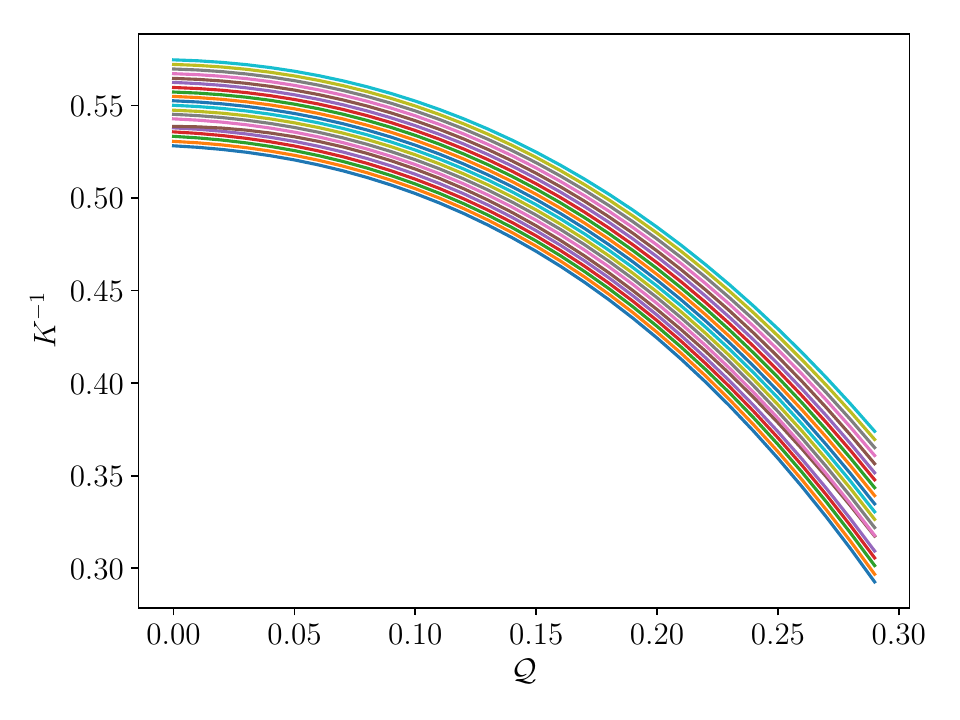}
        \subcaption{$K^{-1}(\mathcal{Q})$ for different values of $\beta$}
    \end{subfigure}
    \begin{subfigure}{0.49\textwidth}
        \includegraphics[width=\textwidth]{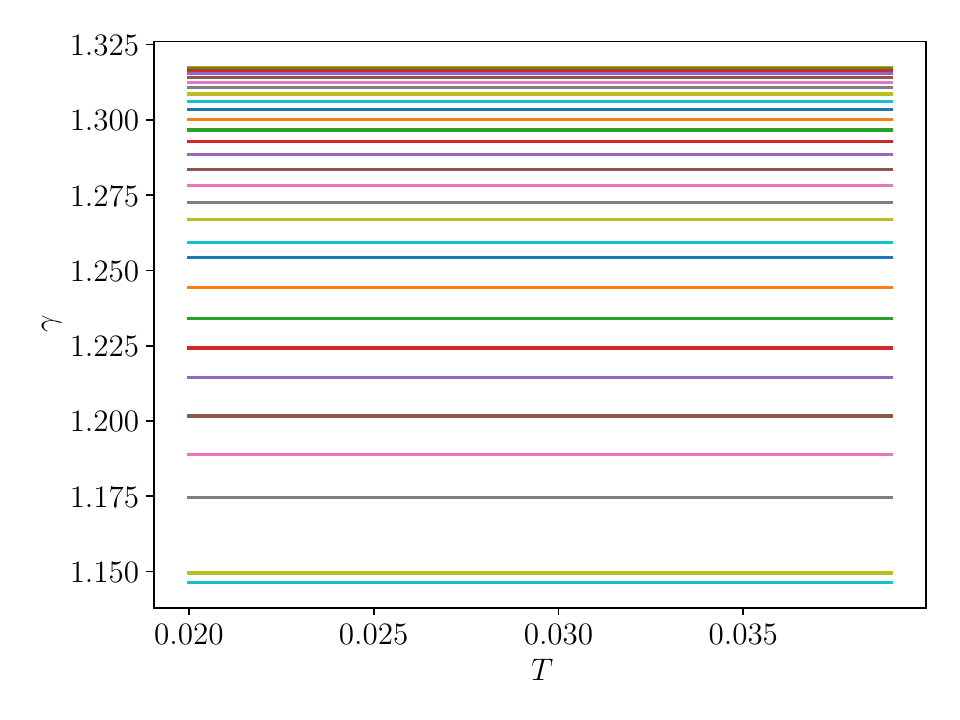}
        \subcaption{$\gamma$ for different values of $\mathcal{Q}$}
    \end{subfigure}
    \caption{Thermodynamic parameters, charge compressibility $K$ and specific heat $\gamma$, from numerical results.}
    \label{fig:K_and_gamma}
\end{figure}

In \cref{fig:free_energy_beta_Q}, we plot constant $\mathcal{Q}$ and $T$ lines of the free energy $F$ and perform a polynomial fit to each curve. Here, it is important to remember that we work in the small $T$ limit and that we only need to perform the match for the gaseous phase, where $\mathcal{Q}\ll 0.5$, since beyond that $S_\text{UV}$ will always be smaller than $S_\text{Gen}$. Thus, while $q$ is not an ideal choice for the expansion parameter, $T$ and $\mathcal{Q}$ are well suited. We use $\mathcal{O}(T^2)$-polynomials to fit the $F(T)$ curves and $\mathcal{O}(\mathcal{Q}^4)$-polynomials for $F(\mathcal{Q})$. We can then take derivatives of those polynomials to obtain $K^{-1}(Q)$ at constant values of $T$ and $\gamma$ for constant values of $\mathcal{Q}$, according to \cref{compressibility,gamma} respectively. This is shown in \cref{fig:K_and_gamma}. Notice that $\gamma$ is constant in $\beta$. This is required by its definition \cite{Davison_2017}
\begin{equation}
    S(T,\mathcal{Q}) = S_0(\mathcal{Q}) + T\gamma(\mathcal{Q}) + \mathcal{O}(T^2).
\end{equation}

\begin{figure}[h]
    \centering
    \begin{subfigure}{0.49\textwidth}
        \includegraphics[width=\textwidth]{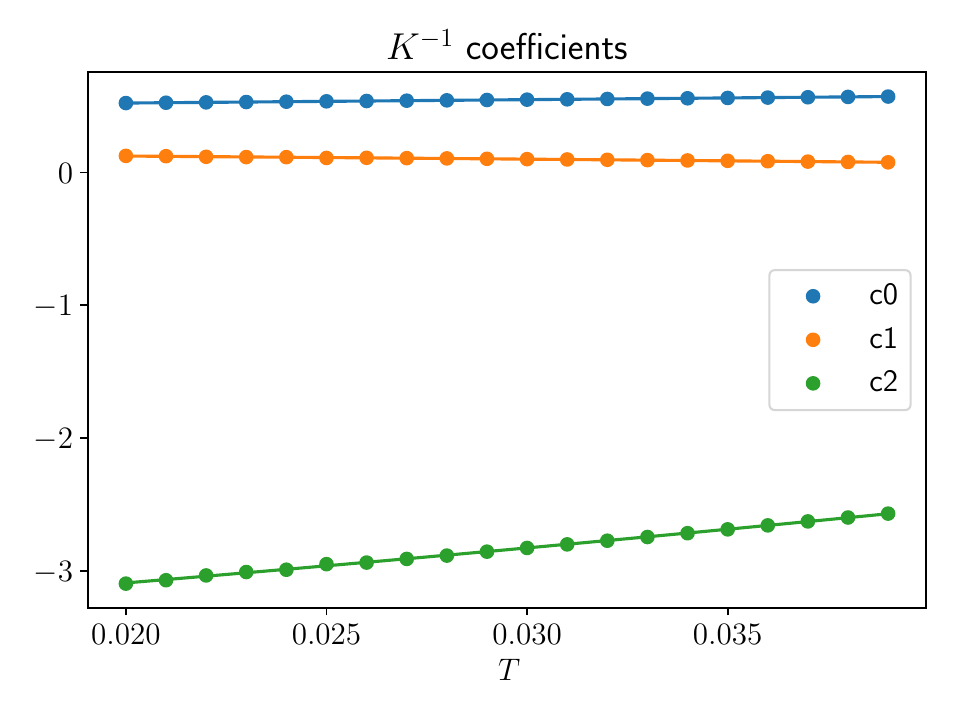}
        \subcaption{Coefficients for $K^{-1}(\mathcal{Q})$ as functions of $T$.}
    \end{subfigure}
    \begin{subfigure}{0.49\textwidth}
        \includegraphics[width=\textwidth]{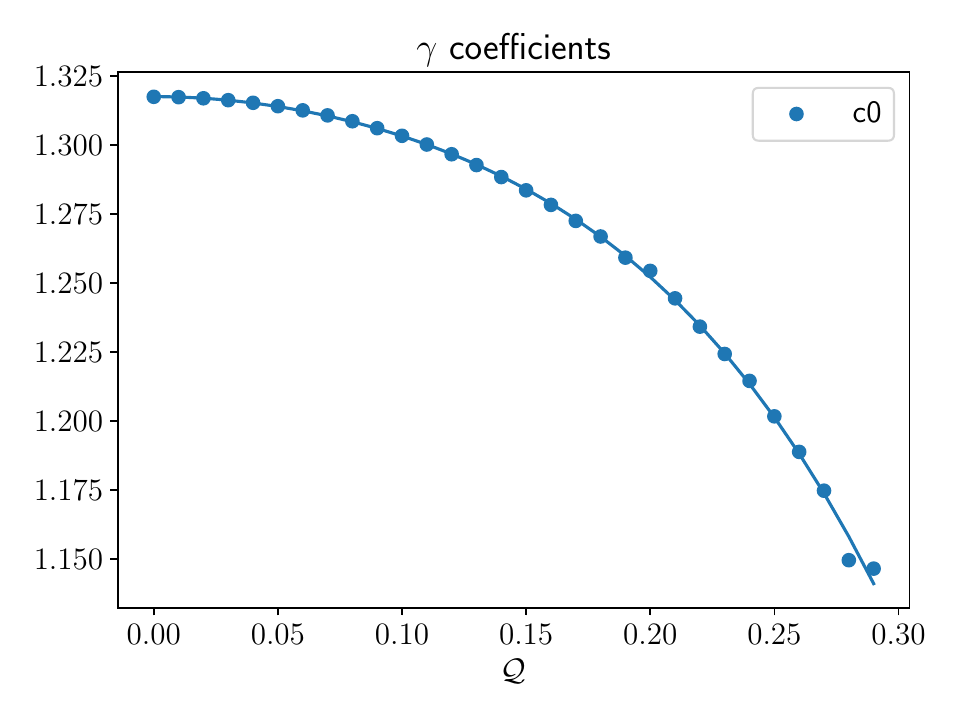}
        \subcaption{$\gamma$ as a function of $\mathcal{Q}$.}
    \end{subfigure}
    \caption{Fit of the polynomial coefficients for the collections of curves $\{K^{-1}_T(\mathcal{Q})\}$ and $\{\gamma_\mathcal{Q}\}$.}
    \label{fig:polynomial_coefs_K_gamma}
\end{figure}

\begin{figure}[ht]
    \centering
    \begin{subfigure}{0.49\textwidth}
        \includegraphics[width=\textwidth]{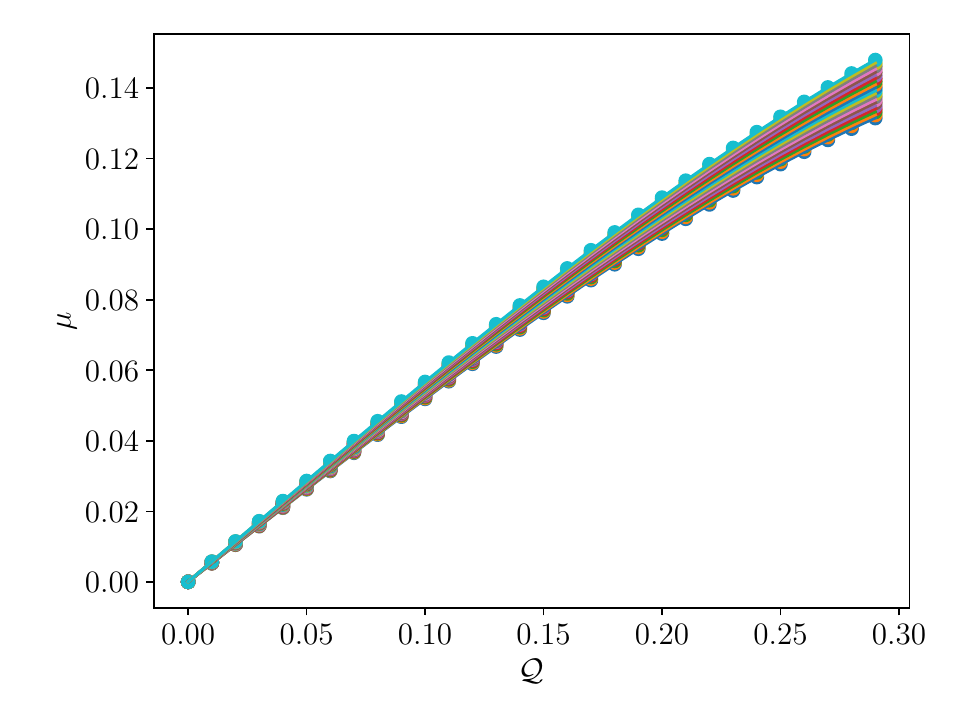}
        \subcaption{$\mu(\mathcal{Q})$ for different values of $T$.}
    \end{subfigure}
    \begin{subfigure}{0.49\textwidth}
        \includegraphics[width=\textwidth]{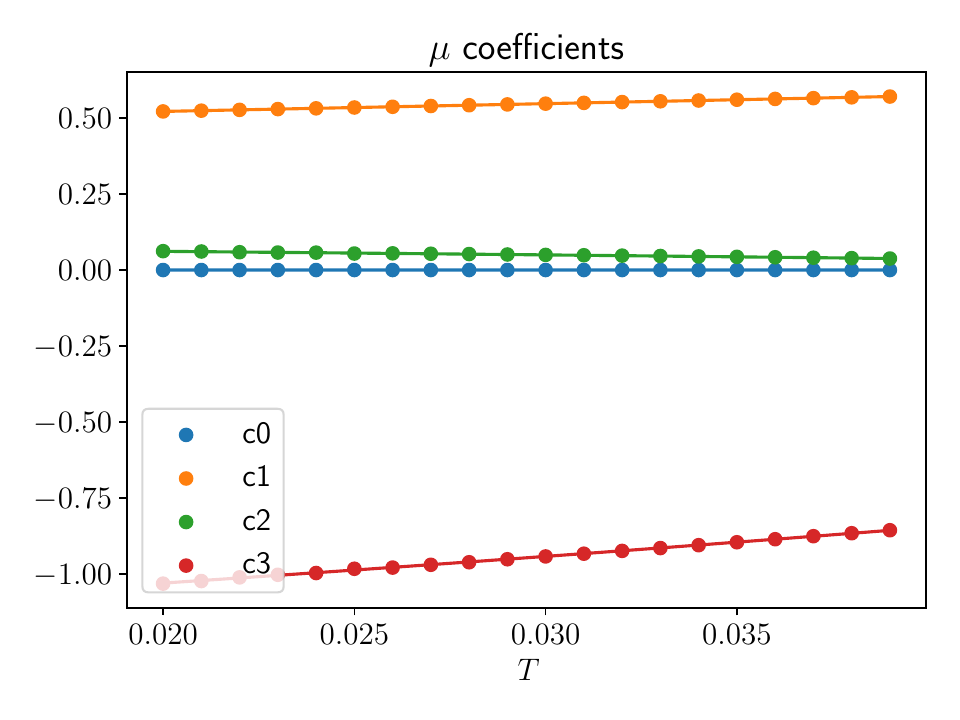}
        \subcaption{Coefficients for $\mu(\mathcal{Q})$ as functions of $T$}
    \end{subfigure}
    \caption{Fit of the polynomial coefficients for the collection of curves $\{\mu_T(\mathcal{Q})\}$.}
    \label{fig:polynomial_coefs_mu}
\end{figure}

We now have a one-parameter family of curves for $K^{-1}(Q)$ and $\gamma$ respectively. However, what we need are explicit expressions for power series expansions in all the arguments for $K^{-1}(T,\mathcal{Q})$ and $\gamma(\mathcal{Q})$. To find the dependence, we plot the polynomial coefficients for the different curves at the respective value of their family parameter and interpolate between them using a polynomial of the respective order given above. This is warranted by the assumption that throughout the gaseous phase $F$ can be described by a smooth polynomial in $T$ and $\mathcal{Q}$. We give the corresponding plots in \cref{fig:polynomial_coefs_K_gamma}. Through an equivalent procedure, we also find a small $T$ and $\mathcal{Q}$ expansion for $\mu(T, \mathcal{Q})$ (see \cref{fig:polynomial_coefs_mu}). Finally, we can give the general expressions for $\mu$, $K^{-1}$ and $T$,
\begin{align}
    \mu(T,\mathcal{Q}) &= (0.5+ 2.5T + 1.9T^2)\mathcal{Q}\nonumber \\
    &\hphantom{=}+(0.1 -0.8T -7.3T^2)\mathcal{Q}^2\nonumber\\
    &\hphantom{=}+(-1.2+ 6.8T+ 39.0T^2)Q^3,\\
    K^{-1}(T,\mathcal{Q}) &=
     0.5 +  2.5T + 1.9T^2 \nonumber\\ 
    &\hphantom{=}+ (0.2 -1.6T -14.6 T^2)\mathcal{Q} \nonumber\\
    &\hphantom{=}+ (-3.5 + 20.5T + 116.9T^2)\mathcal{Q}^2,\\
    \gamma(\mathcal{Q}) &= 1.3  -1.6\mathcal{Q}^2+ 3.0 \mathcal{Q}^3 -16.65 \mathcal{Q}^4,
\end{align}
where the coefficients given here are not the exact ones used for calculations but are rounded to the first decimal place.

\section{R\'{e}nyi-2 vs.\ von Neumann Entropy} \label{sec:neumannvrenyi}
\begin{figure}
    \centering
    \includegraphics[width=\textwidth]{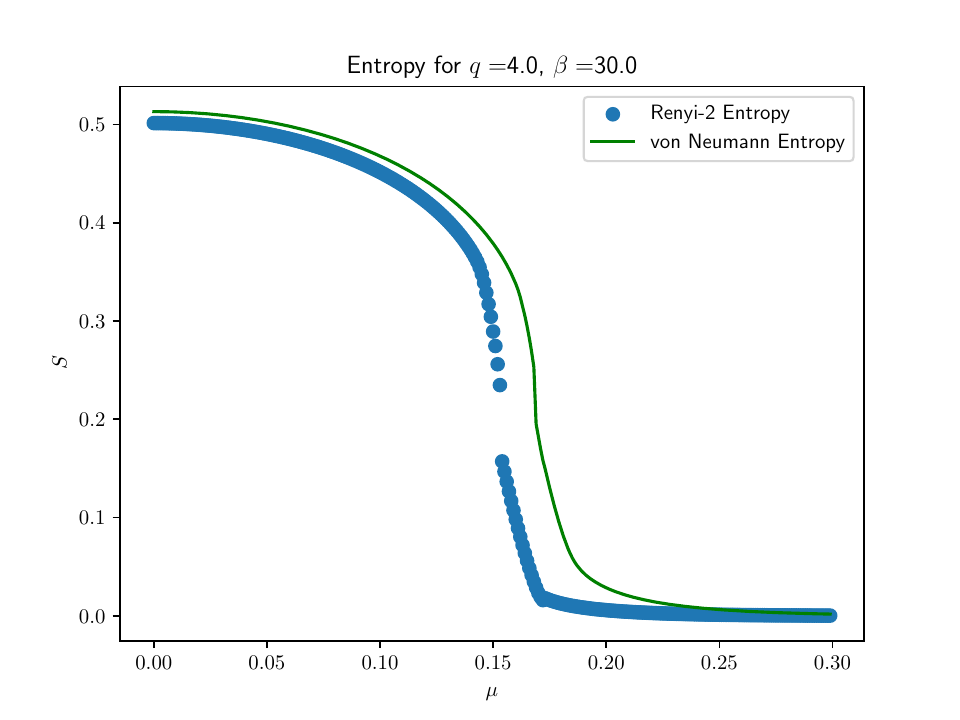}
    \caption{The von Neumann and R\'{e}nyi-2 entropy in the grand canonical ensemble of the complex SYK at $q=4$ and $\beta=30$. Numerical parameters are $\Lambda=2^{17}$ for the discretization and $\epsilon = 10^{-14}$ for the max. error, in the notation of \cite{cao2021thermodynamic}.}
    \label{fig:neumannvrenyi}
\end{figure}

In several places throughout this paper, we use R\'{e}nyi-2 entropy instead of von Neumann entropy to characterize the cSYK TFD after measurement of $M$ fermions. This is a trade off between accuracy and computation time. Calculating the von Neumann entropy numerically in many cases would unnecessarily use up additional computational resources that could instead be used to scan additional portions of the parameter space. In the bulk on the other hand, the situation is quite the opposite. Calculating R\'{e}nyi entropy here would require us to go to a replica version of (nearly) $AdS_2$ \cite{Callebaut_2023,Lewkowycz:2013nqa}, whereas the von Neumann entropy is readily available via the JT on-shell action. We therefore use von Neumann entropy on the JT side. We will argue here that we are nevertheless justified in comparing the two.\\
\\
To show that R\'{e}nyi-2 entropy is indeed a good approximation for the von Neumann entropy, we will solve the cSYK model numerically in the grand canonical ensemble. This is done by employing the algorithm explained in \cite{cao2021thermodynamic} and used in the previous appendices. In the grand canonical ensemble, the von Neumann entropy reduces to the thermodynamic entropy
\begin{equation}
    S_N = (1-\beta\partial_\beta)\log Z \approx (\beta\partial_\beta -1 ) I^{*}.
\end{equation}
With this, we can calculate $S_N$ from our numerical results by computing the on-shell action $I^{*}$ for multiple values in $\beta$ at constant chemical potential $\mu$ and differentiating numerically. To get the R\'{e}nyi-2 entropy, we go to the replica-2 manifold version of our theory. Which here simply amounts to doubling $\beta$ at constant $\mu$. The R\'{e}nyi-2 entropy is thus given by
\begin{equation}
    S_{R2} = 2I^{*}(\beta) - I^{*}(2\beta).
\end{equation}
Figure \ref{fig:neumannvrenyi} shows the result of those two calculations at $q=4$ and $\beta=30$. As we can see, the R\'{e}nyi-2 entropy is a good approximation for the von Neumann entropy on both ends of the curve, although for small $\mu$ it seems to be systematically smaller by a constant amount (we account for this offset in \cref{sec:jt/syk}). However, it deviates substantially from it around the transition point from the gaseous to the liquid phase. 

\section{Micro Canonical Propagator from Full Measurement and Translation Invariance}\label{app: Check for translation invar}
We will show that in the $m=1$ case the propagators defined in \crefrange{eq:G11}{eq:G33} can be expressed in terms of the microcanonical propagator. This is used in \cref{sec:teleportation}. We consider the case where all fermions are measured to be in the positive charge state. This will project the system onto the $Q=1/2$ subsector, which contains a single state $\ket{\uparrow\dots \uparrow} = \ket{1/2,E_{Q_\text{max.}}}$. Let us consider  the $G_{11}$ propagator
\begin{align}
    G_{11}(\tau_1, \tau_2) &= \frac{1}{Z_{m=1}}\frac{1}{2M} \sum_{k=1}^M \braket{T \left[ \left(c_k-c_k^\dagger\right)(\tau_1) \left(c_k-c_k^\dagger\right)(\tau_2) \right]}_m \nonumber\\
    &= \frac{1}{Z_{1}}\frac{1}{2N} \sum_{k=1}^N \bra{\uparrow\dots\uparrow} T\left[ \left(c_k-c_k^\dagger\right)(\tau_1) \left(c_k-c_k^\dagger\right)(\tau_2) \right] e^{-\beta E_{Q_\text{max.}}} \ket{\uparrow\dots\uparrow} \nonumber\\
&= \frac{1}{Z_{1}}\frac{1}{2N} \sum_{k=1}^N \bra{\uparrow\dots\uparrow} T\left[ -c_k^\dagger(\tau_1)c_k(\tau_2) -c_k(\tau_1)c_k^\dagger(\tau_2)\right] e^{-\beta E_{Q_\text{max.}}} \ket{\uparrow\dots\uparrow} .
\end{align}
In the second term we can flip all positive charges to negative charges, and interchange all annihilation with creation operators and vice versa (this is done by a combination of time reversal and charge conjugation, which in combination is a unitary operation)
\begin{align}
     \bra{\uparrow\dots\uparrow} T\left[c_k(\tau_1)c_k^\dagger(\tau_2)\right]  \ket{\uparrow\dots\uparrow} = \bra{\downarrow\dots\downarrow} T\left[c_k^\dagger(\tau_1)c_k(\tau_2)\right]  \ket{\downarrow\dots\downarrow},
\end{align}
which yields
\begin{align}
     G_{11}(\tau_1,\tau_2) &= -\frac{1}{2N} \Tr{\sum_{k=1}^N  T\left[ c_k^\dagger(\tau_1)c_k(\tau_2) \right]}_{E_{Q_\text{max.}}} \\
     &= -\frac{1}{2} \Tr{ T\left[ c_k^\dagger(\tau_1)c_k(\tau_2) \right]}_{E_{Q_\text{max.}}},
\end{align}
where we used that $Z_1=e^{-\beta E_{Q_\text{max.}}}$ and the trace is over the $Q=1/2$ and $Q=-1/2$ state, both of which have the same energy $E\equiv E_{Q_\text{max.}}$. Similarly for $G_{22}$, we find
\begin{align}
   G_{22} (\tau_1,\tau_2) &= \frac{1}{2} \Tr{  T\left[ c_k^\dagger(\tau_1)c_k(\tau_2) \right]}_{E_{Q_\text{max.}}} \\
&= - G_{11} (\tau_1, \tau_2).\label{eq: G22 and G11 realtion}
\end{align}
For $G_{12}$ and $G_{21}$, things are a bit different. We have
\begin{align}
    \begin{split}
     G_{12}(\tau_1,\tau_2) &= 
     \frac{1}{2N} \sum_{k=1}^N  (-\bra{\uparrow\dots\uparrow}T\left[ c_k^\dagger(\tau_1)c_k(\tau_2) \right]\ket{\uparrow\dots\uparrow}\\ &\hphantom{= 
     \frac{1}{2N} \sum_{i=1}^N  (}+ \bra{\downarrow\dots\downarrow}T\left[ c_k^\dagger(\tau_1)c_k(\tau_2) \right]\ket{\downarrow\dots\downarrow} )
    \end{split}\\
    &= -G_{21}(\tau_1,\tau_2),
\end{align}
which we can rewrite as
\begin{align}
    \begin{split}
     G_{12}(\tau_1,\tau_2) &= 
     \frac{1}{N} \sum_{k=1}^N  (-\bra{\uparrow\dots\uparrow}T\left[ c_k^\dagger(\tau_1)c_k(\tau_2) \right]Q\ket{\uparrow\dots\uparrow}\\
&\hphantom{= 
     \frac{1}{N} \sum_{i=1}^N  (}- \bra{\downarrow\dots\downarrow}T\left[ c_k^\dagger(\tau_1)c_k(\tau_2) \right]Q\ket{\downarrow\dots\downarrow} )
    \end{split} \\
&= - \Tr{  T\left[ c_k^\dagger(\tau_1)c_k(\tau_2) \right]Q}_{E_{Q_\text{max.}}}\label{eq: G11Q in operators}\\
&= -G_{21}(\tau_1,\tau_2).
\end{align}
Consequently as operator equations, the following relations hold 
\begin{align}
    G_{12}(\tau_1,\tau_2) &= 2G_{11}(\tau_1,\tau_2)Q,\label{eq:G12 and G11 realtion}\\
    G_{21}(\tau_1,\tau_2) &= 2G_{22}(\tau_1,\tau_2)Q.\label{eq: G21 and G11 realtion}
\end{align}
Finally, since there are no unmeasured fermions, $G_{33}$ is ill-defined. For consistency, we will set it to zero.

Additionally, we check for time-translation symmetry when $m=1$ as this will greatly simplify the calculations for solving the Liouville equation \eqref{eq: Lioville eq for g} (i.e.\ the teleportation protocol is applied by measuring the full left side). Consider
\begin{equation}
    E_m = \big\langle H\big\rangle_m = \big\langle H\ketbra{L_M} \big\rangle.
\end{equation}
Via the Ehrenfest theorem we then have,
\begin{equation}\label{eq: alt constraint for translation invar}
\frac{\text{d}E_m}{\text{d}\tau} = \braket{[H,H]\ketbra{L_M}{L_M}} + \braket{H,[H,\ketbra{L_M}{L_M}]}    
\end{equation}
The first term in the right side of \eqref{eq: alt constraint for translation invar} is always zero while the second term is only zero if either all or none of the fermions have been measured i.e.\ when $m\in\{0,1\}$ (since then $\ket{L_M}$ is an eigenstate to $H$). Hence, using Noether's theorem we see that we only have translation invariance if $m\in\{0,1\}$.
\newpage
\bibliographystyle{utphys.bst}
\bibliography{literature}
\end{document}